\documentclass[%
 aps,
 pre,%
 amsmath,amssymb,
 reprint,%
]{revtex4-2}

\usepackage{graphicx}
\usepackage{epstopdf}
\usepackage{bm}
\usepackage{stmaryrd}
\usepackage{natbib}
\usepackage{xcolor}
\usepackage{hyperref}
\hypersetup{
  colorlinks   = true, 
  urlcolor     = blue, 
  linkcolor    = blue, 
  citecolor   = blue 
}
\newcommand{\orcidlink}[1]{\href{https://orcid.org/#1}{\includegraphics[width=10pt]{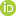}}}

\begin{document}

\title{Turbulence Generation by Shock Interaction with a Highly Non-Uniform Medium}
\author{Seth Davidovits\orcidlink{0000-0002-4808-7286}$^{1}$}
\thanks{davidovits1@llnl.gov}
\author{Christoph Federrath\orcidlink{0000-0002-0706-2306}$^{2,3}$}
\author{Romain Teyssier\orcidlink{0000-0001-7689-0933}$^{4}$}
\author{Kumar S. Raman$^{1}$}
\author{David C. Collins\orcidlink{0000-0001-6661-2243}$^{5}$}
\author{Sabrina R. Nagel\orcidlink{0000-0002-7768-6819}$^{1}$}
\affiliation{$^{1}$Lawrence Livermore National Laboratory, Livermore, California 94550, USA}
\affiliation{$^{2}$Research School of Astronomy and Astrophysics, Australian National University, Canberra, ACT 2611, Australia}
\affiliation{$^{3}$Australian Research Council Centre of Excellence in All Sky Astrophysics (ASTRO3D), Canberra, ACT 2611, Australia}
\affiliation{$^{4}$Department of Astrophysical Sciences, Princeton University, Princeton, New Jersey 08544, USA}
\affiliation{$^{5}$Department of Physics, Florida State University, Tallahassee, FL 32306, USA}

\begin{abstract}
An initially planar shock wave propagating into a medium of non-uniform density will be perturbed, leading to the generation of post-shock velocity perturbations. Using numerical simulations we study this phenomenon in the case of highly-non-uniform density (order-unity normalized variance, $\sigma_{\rho}/\overline{\rho} \sim 1$) and strong shocks (shock Mach numbers $\overline{M}_s \gtrsim 10$). This leads to a highly disrupted shock and a turbulent post-shock flow. We simulate this interaction for a range of shock drives and initial density configurations meant to mimic those which might be presently achieved in experiments. Theoretical considerations lead to scaling relations, which are found to reasonably predict the post-shock turbulence properties. The turbulent velocity dispersion and turbulent Mach number are found to depend on the pre-shock density dispersion and shock speed in a manner consistent with the linear Richtymer-Meshkov instability prediction. We also show a dependence of the turbulence generation on the scale of density perturbations. The post-shock pressure and density, which can be substantially reduced relative to the unperturbed case, are found to be reasonably predicted by a simplified analysis that treats the extended shock transition region as a single normal shock.
\end{abstract}

\maketitle

\section{Introduction}

Shock propagation through a field of non-uniform density is a problem of interest in a variety of settings where shocks arise, including aeronautics (e.g. \citenum{moore1954,hesselink1988}), astrophysics (e.g. \citenum{giacalone2007,inoue2013,banda-barragan2021}), and inertial confinement fusion (e.g. \citenum{velikovich2007,velikovich2012,ali2018}). The problem has a significant conceptual overlap with the problem of shock-turbulence interaction, where a shock propagates through turbulent flow. However, in the turbulence case, the pre-shock field may or may not have density perturbations, depending on the assumed compressibility of the initial turbulence.

The motivating application for the present work is the controlled generation of turbulence for laboratory study, as may be useful for laboratory astrophysics\citep{remington2006} or other applications. For example, compressible turbulence plays a key role in the formation of stars\citep{federrath2018}.  A variety of setups for generating such turbulence exist, often utilizing the collision of laser ablated plasmas to generate and study turbulence (e.g. \citenum{meinecke2015,tzeferacos2018,white2019}). 
Here we consider an approach based on the interaction of a strong shock with a medium of non-uniform density as might be realized in a laser-driven high-energy-density (HED) shock tube\citep{nagel2017}. Various fabrication techniques may permit the specification of the pre-shock density field in such a shock tube\citep{hamilton2016,saha2018}. Here we consider an idealized, yet close to ``realizable'', version of such a shock tube. This setup also has similarities to a series of experiments studying mixing processes in inertial confinement fusion\citep{murphy2016,haines2020}.

Various theoretical treatments have been developed to predict the post-shock state when a shock propagates through a non-uniform medium, particularly when one broadens to consider shock-turbulence interaction \citep{moore1954,ribner1955,ribner1987,durbin1992,mahesh1995,lele1992,zank2002,velikovich2007,huete2011,velikovich2012,kitamura2016,chen2019}.  Most of these treatments are, at least formally, only applicable in the case where the upstream (pre-shock) perturbations are small in some sense and material can only pass through the shock front once in the interaction. We note that the `quasi-equilibrium' theory of \citet{donzis2012}, which is also formally applicable for small upstream perturbations (small in $K \equiv M_t/R_{\lambda}^{1/2} (\overline{M}_s - 1)$, with $M_t$ and $R_{\lambda}$ the upstream turbulent Mach number and Reynolds number, respectively), has been applied successfully to interactions where local `holes' are formed in the shock due to pre-shock non-uniformity\citep{chen2019}. Such a theory might also be applied to the case with only pre-shock density perturbations, particularly if these perturbations are sufficiently smooth. When there are only upstream density perturbations (that is, as we consider here, when there are no upstream velocity perturbations), the smallness of upstream perturbations is generally in some measure of the density perturbations relative to the mean density, with the precise definition varying between different prior treatments. 

Here we characterize the pre-shock density non-uniformity by its variance $\sigma_{\rho}$, normalized to the mean density, $\overline{\rho}$, and consider cases where $\sigma_{\rho}/\overline{\rho} \sim 1$, violating the small-perturbation assumptions of most theoretical approaches. Further, with these large pre-shock density variations, the single shock front can break into multiple fronts. As such, our approach here is to conduct a suite of numerical simulations of an (initially planar) shock propagating through an extended region of non-uniform density. Ultimately we would like to be able to predict the post-shock turbulent state given the controllable parameters, such as the shock drive and the features of the pre-shock density non-uniformity.

Analyzing the simulations, we find the post-shock state can be reasonably predicted. In particular, as observed by \citet{inoue2013}, we find here that the scaling of the post-shock (turbulent) velocity dispersion with the shock speed and with the relative pre-shock density variance is reasonably predicted by adapting the linear analysis of the Richtmyer-Meshkov instability (RMI)\citep{richtmyer1960}. Expanding on this result, we show that the scale of pre-shock density features influences the turbulence as well. In the present cases, the pre-shock density non-uniformity has a dominant perturbation scale, which influences the post-shock turbulence generation as an effective driving scale.

We also show predictability for a post-shock volume-averaged turbulent Mach number. This turbulent Mach number is primarily influenced by the pre-shock density structure, with the response to the drive (shock velocity) saturating for strong shocks. 

Although the present cases often have an extended shock transition region with a broken-up shock front, we test a simplified analysis which treats the shock front as a single (average) normal shock. For the present suite of simulations, we find this approach is reasonably successful in explaining the post-shock pressure and density behaviors. In the most perturbed shocks here, the post-shock pressure can be less than half its unperturbed value. This theoretical treatment suggests that the post-shock pressure is relatively insensitive to the anisotropy of the turbulence generated, while the post-shock density is rather sensitive to anisotropy, especially in the 3D case.

The paper is laid out as follows. Section \ref{sec:approach} describes the setup for our study and simulations; additional details on this are also contained in the Appendices \ref{sec:A}, \ref{sec:B}. The primary results are given in Sec. \ref{sec:results}. We provide further discussion of the results in Sec. \ref{sec:discussion} and then summarize in Sec. \ref{sec:summary}.

\section{Approach} \label{sec:approach}

\begin{figure}
\includegraphics[width=\columnwidth]{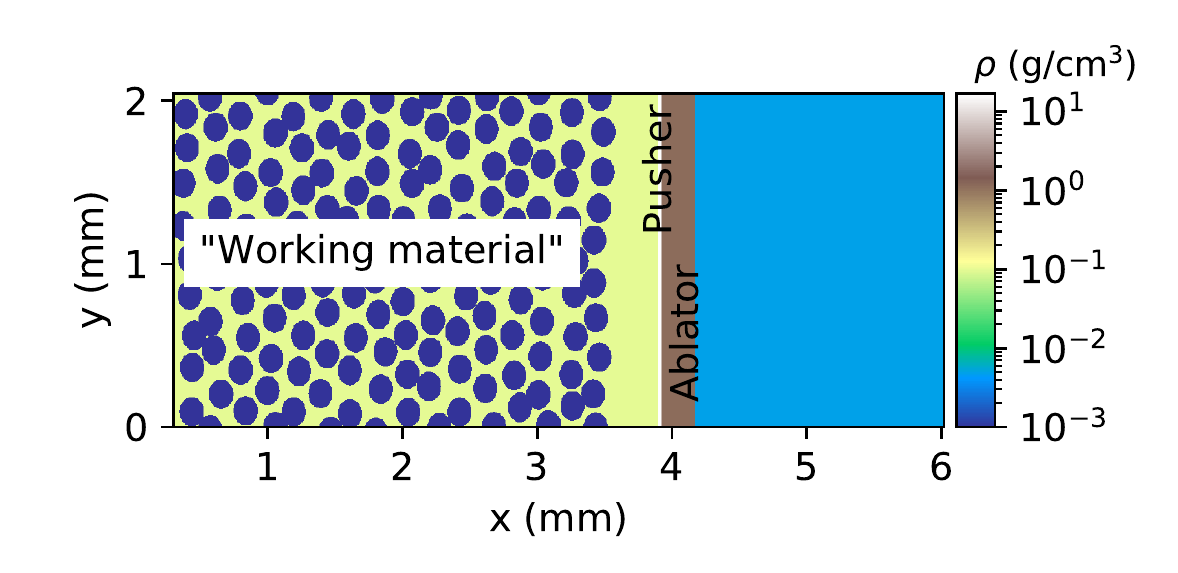}
\caption{\label{fig:sketch} An example initial condition for a 2D simulation to illustrate the setup of the present simulations. A radiation source from the right-hand-side boundary acts on the ablator to launch a shock that travels to the left through a thin, dense, pusher (white vertical strip immediately to the left of the ablator) and then through the ``working material''. In the working material, which has a non-uniform density owing to the presence of low-density ``voids'', turbulence is generated by the shock passage. }
\end{figure}

\subsection{Setup and simulation description} \label{sec:setup}
Although shock interaction with highly non-uniform density arises in a variety of settings, the underlying motivation here is to use this mechanism to generate turbulence for laboratory study. This guides the present setup, in which we simulate an idealized high-energy-density (HED) shock tube (e.g. \citet{nagel2017}); an example initial condition is shown in Fig.~\ref{fig:sketch}. In this setup, a radiation source from the right-hand-side (RHS) ablates a (plastic) sheet (the ablator), launching a shock which passes through a dense pusher and into the non-uniform-density medium of interest (the ``working material''). The shock runs through the working material well ahead of the pusher, and the analysis occurs before the shock exits the working material. In the present work, our focus is on the turbulence generation dynamics in the working material, that is, on the basic problem of a shock interacting with non-uniform density. Once the shock enters into the working material, we turn off radiation physics and the problem becomes a purely hydrodynamic one. 

Our simulations use the radiation-hydrodynamics code HYDRA\citep{marinak2001}, but, to simplify theoretical comparisons and to simplify comparison with shock-turbulence work, we run hydrodynamic-only simulations (once the shock profile is established) and treat the working material as an ideal monatomic ($\gamma=5/3$) gas. We also ignore thermal conduction in the working material; while it is common in shock-turbulence interaction studies to include thermal conduction (e.g., treating the material as air), analytic predictions for the post-shock flow typically do not account for it. The present simulations utilize artificial viscosity to capture the shock. We also neglect explicit (physical) viscosity, so that the simulations should not be regarded as direct numerical simulation (DNS), but are rather in the spirit of implicit large eddy simulations\citep{sagaut2006}. Here we focus our analysis on large-scale quantities (e.g., the root-mean-square turbulent velocity and average density, pressure, or temperature), which are expected to be least influenced by the (numerical) dissipation details (see, e.g., Ref. \citenum{kitsionas2009}). As described later, we test sensitivity of the present results to the simulation resolution, and believe the conclusions of the work are robust in this regard.

As a result of the idealities just described, while we choose to report lengths and certain other quantities in HED-shock-tube relevant units, we also provide the information necessary to non-dimensionalize quantities in terms of the pre-shock conditions for comparison with other idealized simulations. Apart from influencing the choices of the size of density non-uniformities relative to the simulation domain, the primary HED-shock-tube feature of the present simulations is the details of the shock drive, which generates a decaying shock and is discussed briefly below and in detail in Appendix \ref{sec:B}.

\begin{table}
\begin{tabular}{ l | r}
\hline
Parameter &  Values (2D) \\
\hline
Shock Mach number [$\overline{M}_s$] & 12, 15, 18, 20, 37 \\
Pre-shock density variance [$(\sigma_\rho/\overline{\rho})_0$] & 0.38, 1.05, 1.9 \\
``Void'' radius [$R$ (microns)] & 23, 45, 90 \\
\hline
\end{tabular}
\caption{\label{tab:parameters} Table of parameters which influence the turbulence generated by the shock and are varied in the present simulations. The 2D simulations cover most of the 45 ($5\times3\times3$) possible parameter combinations, and also include repeats of certain combinations. These repeats use different random seeding for ``void'' placement to test variability in the analysis (see text). There are six 3D simulations, which have approximate parameter values, listed in the format [$\overline{M}_s$,$(\sigma_\rho/\overline{\rho})_0$,$R$] of: [18,0.3,120], [18,0.9,120], [18,1.07,120], [37, 0.9, 120], [18, 0.9, 90], [37, 0.9, 90].}
\end{table}

As described above, an initially planar shock is launched into the RHS of the working material, which has a short ($\sim0.2$ mm) uniform section before the non-uniformity begins. The shock propagates to the left (negative $x$ direction) for a length ($\sim 2.5$ mm) that is substantially larger than the scales of the non-uniformity ($\lesssim 0.2$ mm, see the ``voids'' in Fig.~\ref{fig:sketch}). After the shock (on average) has propagated this length (to an average position of $x \approx 1$ mm in Fig.~\ref{fig:sketch}), the post-shock quantities are analyzed; here we largely report on averaged quantities, which are averaged over a post-shock strip of fixed width (0.2 mm) a distance ($\sim$ 0.1 mm) behind the end of the shock transition. The choice of averaging window is discussed in more detail in Sec. \ref{sec:averaging} below. Figure \ref{fig:fig_snapshot} (a) shows an example density snapshot when the shock has propagated this distance, for a two-dimensional (2D) simulation. Also shown in Fig. \ref{fig:fig_snapshot} are the 1D profiles (averaged in $y$), which we discuss later, as well as indicators for the analysis region.

\begin{figure}
\includegraphics[width=\columnwidth]{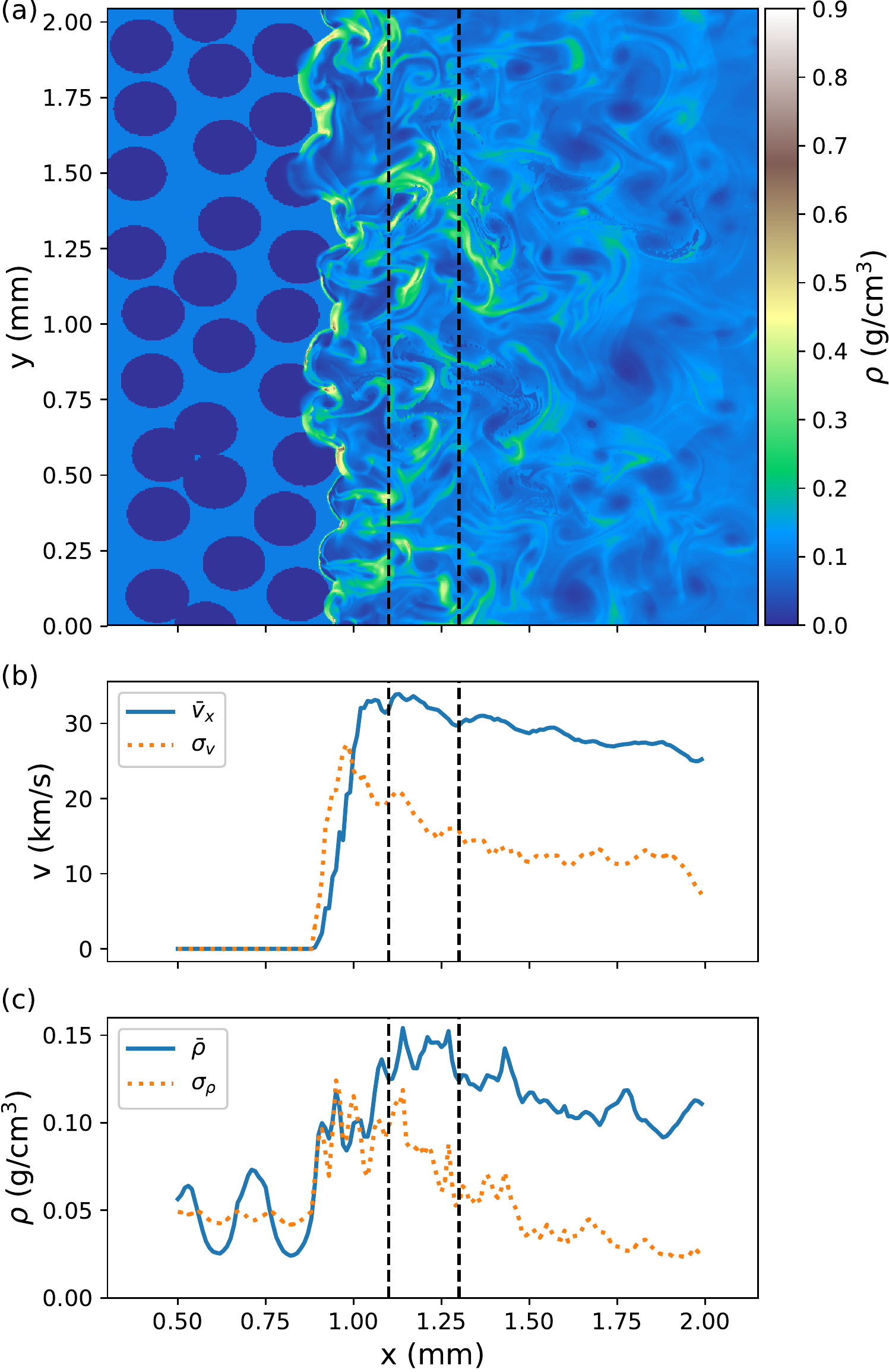}
\caption{\label{fig:fig_snapshot} (a) A density plot from a typical 2D simulation around the time of analysis, after the (average) shock front has propagated $\sim$2.5 mm through the non-uniform material to $x \sim 1$ mm. Vertical dashed lines indicate the analysis region, and the density is shown in units of g/cm$^3$. Only the ``working material'' region is plotted, the pusher and ablator material (see Fig. \ref{fig:sketch}) are out of frame at $x \gtrsim 2.25$ mm. Plots (b) and (c) show, respectively, 1D profiles (averaged over $y$) for the mean $x$ directed velocity ($\overline{v}_x$) or the (turbulent) velocity dispersion ($\sigma_v$) and the mean density $\overline{\rho}$ or the density dispersion ($\sigma_\rho$).}
\end{figure}

For this study, we conduct a suite of 2D simulations, supplemented with a smaller set of 3D simulations. This permits us to study the effects of varying the pre-shock density structure as well as the (average) shock Mach number (drive strength), while also enabling some comparison to 3D. For the 2D simulations, we vary across five different initial shock strengths (Mach numbers), and, for the pre-shock density non-uniformity, three different density-variance values and three different scale parameters. Each of these three parameters (shock velocity, pre-shock density variance, scale of density non-uniformity) influences the magnitude of the turbulent velocity generated post shock. The parameter values, which we now discuss, are summarized in Tab. \ref{tab:parameters}.

Here we characterize the pre-shock material by its mean density ($\overline{\rho}_0 \approx 0.05$  g/cm$^3$ for all cases) and its (normalized) density variance, reported as the standard deviation of the density divided by the mean density $(\sigma_\rho/\overline{\rho})_0$. These quantities are computed over the ``working material'', but excluding the uniform region on its right side (see Fig.~\ref{fig:sketch}). The three density-variance conditions we use have $(\sigma_\rho/\overline{\rho})_0 \sim 0.38,\,1.05,\,1.9$, which are created by embedding circles (2D) or spheres (3D) of lower or higher density in a uniform-density background. Figures \ref{fig:sketch} and \ref{fig:fig_snapshot} show examples of the cases with $(\sigma_\rho/\overline{\rho})_0 \sim 1.05$, which have low-density (10$^{-3}$ g/cm$^3$) circular ``voids'' covering approximately half the area in the initial pre-shock material, which has a density of 0.1 g/cm$^3$ in the non-void regions (resulting in $\overline{\rho}_0 \approx 0.05$). The scale of the pre-shock density modulations is controlled by the radius of these ``voids"; in 2D we use $R \approx 23,\,45,\,90$ microns, while for 3D we run cases with $R \approx 90,\:120$ microns. The $(\sigma_\rho/\overline{\rho})_0$ values for the 3D simulations vary somewhat from these 2D values. More details on the ``working material'' initial conditions for the different cases are described in Appendix \ref{sec:A}.

The five shock strengths are characterized by their averaged Mach number ($\overline{M}_s \sim 12,\,15,\,18,\,20,\,37$; in our units the pre-shock sound speed is $C_{s,0} \approx 2.4$ km/s and all materials are initialized at room temperature, 293 K). Since the initial temperature in the ``working'' material is uniform, but the density is non-uniform, the initial (ideal gas) pressure will be non-uniform. While in a real experiment initial material strength could keep this initial condition stationary, here we use a so-called ``quiet start'' routine that prevents premature motion, with a temperature-based shutoff threshold well below the post-shock temperature in our weakest shock.

In the absence of any pre-shock density perturbations, the shock strength in each case decays as it traverses the ``working" material (it also decays when there are density perturbations). Although the detailed flow conditions set up by the ablation which launches the shock are complicated, this decay is not unexpected given there is a finite reservoir of high pressure available to drive the shock (the shocks launched in these laser-driven HED shock tubes are more akin to blast waves than constant-pressure shocks). As a result, the shock velocity also decays somewhat and the average values reported here come from a linear fit to the position versus time of the average shock position as it propagates in the vicinity ($\sim \pm 0.5$ mm) of the ``observation" position. The position of the shock at any given time is defined by the maximum of the gradient of the averaged (1D) velocity $\overline{v}_x (x)$ in the shock transition region. In Fig. \ref{fig:fig_snapshot} (b) and (c), respectively, the decaying shock strength can be seen in the (gradually) decreasing mean velocity ($\overline{v}_x$) and average density ($\overline{\rho}$) in the post-shock region. The (turbulent) velocity dispersion, $\sigma_v$, also decreases behind the shock (moving to larger $x$); both the decaying underlying shock profile and dissipation (turbulent decay) can contribute to this decrease.

The simulations for all cases use an initially uniform grid. For the 2D runs the grid in the ``working" material consists of either $\sim$846x512 cells ($R \approx 45$, $R \approx 90$) or $\sim$1694x1024 cells ($R \approx 23$), where the initial domain size is $\sim$3.8 by 2.048 ($x$ by $y$, in mm). The 3D runs use either $\sim$492x300x300 ($R \approx 120$) or  $\sim$657x400x400 ($R \approx 90$) cells in a domain that is $\sim$3.8 by 2.048 by 2.048 mm. Our simulations use HYDRA's arbitrary Lagrange Eulerian (ALE) hydrodynamics features, so that the mesh tends to move with the flow. This results naturally in a more closely spaced post-shock mesh, where the flow scales are also finer. In addition to the runs shown, we have rerun a subset of the 2D cases with double the resolution in each dimension in order to test numerical convergence; these runs do not appear to change the scalings for averaged quantities we present here. As an example, running the case $R\approx45$, $(\sigma_\rho/\overline{\rho})_0 \sim 1.05$, $\overline{M}_{s} \sim 18$ with double the resolution is observed to result in changes to analysis-region averaged (see below) quantities of between  0.6\% - 2.9\%, depending on the quantity. Measured in terms of initial cells per void, the $R \approx 23, \, 45$ cases are the lowest effective resolution.

The boundary conditions in the shock-normal directions ($y$, $z$) are for zero normal velocity (zero velocity into the boundary). Since the material is stationary for $x < x_{shock,minimum}$, the boundary on this side has no effect up to the time of the analysis. In a similar way, the shock propagates well ahead of influences from the boundary at $x_{maximum}$, see Appendix \ref{sec:B} for more description of the drive side.

\subsection{Generation of analysis quantities}\label{sec:averaging}

Here we describe the generation of post-shock analysis quantities in this work. First we discuss the choice of analysis (averaging) window, which induces some scatter in the analysis. Then, we discuss the averaging procedure within this window.

In general, we are interested in the generation of post-shock turbulence in the interaction of a shock with a non-uniform medium, in a setup similar to what might be achieved experimentally. For the current setup, all post-shock quantities may vary spatially in the shock-normal ($x$) direction; on top of any such variation, we also have the (statistical) variations caused by the compressible turbulence. Further, it is not practical to collect significant statistics on the large number of simulation cases here. As a result, we choose to average over a finite region of the post-shock to provide some statistics, which represents a choice in determining the ``post-shock'' quantities (with analogy to experiment).

Drawing a comparison to shock turbulence interaction studies, even in cases with converged statistics, there are different ways to define post-shock quantities, for example, the peak in the post-shock after an adjustment period, or by some sort of extrapolation back to the average shock position (see, e.g., the discussion in \citet{larsson2013}). Here, we simply define the post-shock quantities by averaging over a window in the post-shock. The position of this window is set to be on order of the large eddy scales behind the shock transition to allow development (this is also the scale of the adjustment region in the shock turbulence case\citep{larsson2013}). The width of the window, also on this order, helps to provide some statistics. Note that we aim not to move the averaging window too far downstream, since we will then necessarily be increasingly conflating the generation of post-shock state with its decay.

\begin{figure*}
\includegraphics[width=\textwidth]{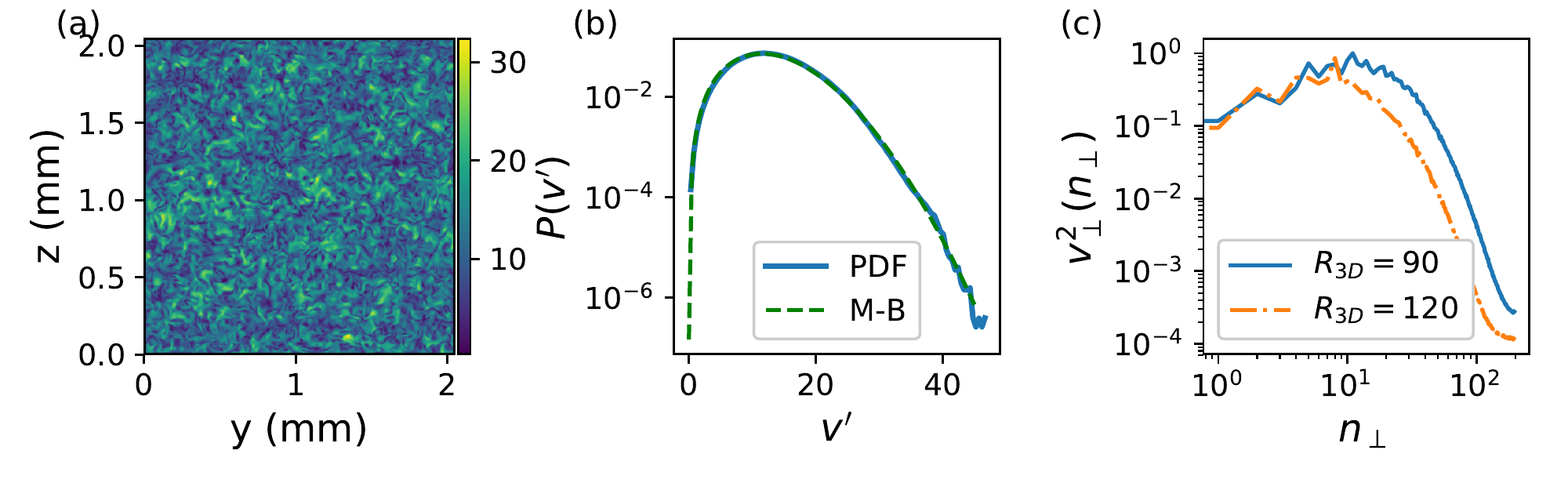}
\caption{\label{fig:fig_3D} Panel (a) shows a slice of the local fluctuating velocity magnitude, $v' = (v'^{2}_{x} + v'^{2}_{y} + v'^{2}_{z})^{1/2}$, (in km/s) through the center of the analysis region ($x=1.2$ mm), for a 3D simulation with $R=90$ and $(\sigma_\rho/\overline{\rho})_0 \sim 0.9$, $\overline{M}_s \sim 18$. This slice then is perpendicular to the shock propagation direction (in a plane parallel to the average shock front). Panel (b) shows the observed (blue solid line) probability distribution function for $v'$ in the analysis region, and also shows (green dashed line) a Maxwell-Boltzmann distribution characterized by the analysis-region-averaged velocity dispersion $\langle\sigma_v\rangle$, $P(v') = 4 \pi v'^2 \mathrm{Exp}[-3 v'^{2}/2 \langle \sigma_v \rangle^2]/(2 \pi \langle \sigma_v \rangle^2/3)^{3/2}$. This is the velocity distribution observed for fully-developed turbulent flows, even at high Mach number (see Fig. A1 in Ref. \citenum{federrath2013}).  Panel (c) shows spectra for the perpendicular piece of the fluctuating velocity ($v_{\perp}^2 = v'^{2}_{y} + v'^{2}_{z}$) as a function of the perpendicular mode number ($n_{\perp} \propto k_{\perp} = (k_{y}^2 + k_{z}^2)^{1/2}$) for both the $R=90$ case and a case with $R=120$ that is otherwise identical. The spectrum in each case is averaged (in $x$) over the analysis region and its magnitude is normalized to a unit peak. We see that in both cases a broad spectrum has developed, with a peak that can be approximately associated with the initial void diameter, $n_{peak,90}=11 \sim 2/(2\times0.09) \sim L_{y,z}/2R$, $n_{peak,120}=8 \sim 2/(2\times0.12)$. Note that the simulation resolution for the $R=120$ case is lower (see Sec. \ref{sec:setup}), and both cases were linearly interpolated onto an identical uniform grid for plotting and computing spectra (the simulations use an ALE mesh, also see Sec. \ref{sec:setup}).}
\end{figure*}

In the case of the turbulent velocity, the choice in defining the ``post-shock'' value is not so consequential for the results presented in Sec. \ref{sec:results}; it is already useful to predict the post-shock turbulence in some specified region downstream of the shock (and we may then work to anticipate its evolution from there). The same can be said for other quantities. In the latter part of the results (Sec. \ref{sec:results}), we compare analytic theory for a single average shock jump to the post-shock density and pressure. Here, the present choice for the post-shock state can be expected to create noise in the comparison. In all figures containing quantities averaged in the post-shock, we display uncertainty bars (vertical and/or horizontal lines) for data points. These are found by redoing the analysis, shifting the center of the averaging window by $\Delta x=\pm0.05$ mm (25\% of its width) upstream or downstream from the nominal position shown in Fig. \ref{fig:fig_snapshot} and described in Sec. \ref{sec:setup}.

We now describe the averaging approach in the analysis window. Averaging is carried out in two stages, in order to separate averaging (first) over the statistically homogeneous directions ($y$ or $y,z$) from averaging (second) over the analysis region, over which quantities should not be, strictly speaking, statistically homogeneous. We now describe the first stage, with the second stage described shortly thereafter. 

In the first stage, quantities are averaged over the shock-normal directions ($y$ or $y$, $z$) to create average profiles that depend on the shock-parallel coordinate (e.g., the profiles shown in (b) and (c) of Fig.~\ref{fig:fig_snapshot}). For a general quantity $f$, the average value in an extent of width $w$, centered on the point $x_i$, is
\begin{align}
\overline{f}\left(x_i;w\right) &= \sum_{x_j \in C_j} V_j f_j / \sum_{x_j \in C_j} V_j \label{eq:bar} \\  
C_j &\equiv [x_i - w/2,x_i + w/2] \nonumber
\end{align}
We define the interval $C_j$ so that the sum over the index $j$ is carried out over all simulation points with $x$ coordinate in the targeted window. As a result of ALE, the initially-uniform mesh is in general non-uniform after shock passage. Thus, this (volume) average is weighted by the cell volume, $V_j$. We pick an averaging window $w = 0.01$ mm (for all cases) that is similar to, but always at least modestly larger than, the initial cell size, such that this first average approximates an average over the statistically homogeneous (symmetry) directions.

In addition to the volume-weighted average defined in Eq.~(\ref{eq:bar}), indicated by the overline, we also consider a density-weighted (Favre) average, $\tilde{f}$. The density-weighted profile $\tilde{f}(x_i;w)$ is defined identically to Eq.~(\ref{eq:bar}), but with the cell mass $m_j$ substituted for the cell volume $V_j$.

Defining difference quantities $f' = f - \overline{f}$ and $f''= f - \tilde{f}$, the volume-weighted dispersion of quantity $f$ is $\sigma_f  = (\overline{f'^2})^{1/2}$. In particular we study in Sec.~\ref{sec:results} the total velocity dispersion, $\sigma_v = (\sigma_{v_x}^2 + \sigma_{v_y}^2 + \sigma_{v_z}^2)^{1/2}$. There we also consider the (density-weighted) Reynolds stresses, $\tau_{\alpha \beta} = \widetilde{v''_\alpha v''_\beta}$.

For most of the results discussed in Sec.~\ref{sec:results}, we go a step further than this first average, by averaging quantities again, this time over the analysis region described above, $x \in [1.1,1.3]$ mm. This second average is indicated by $\langle \rangle$, and is either volume-weighted or density-weighted to match the inner averaging. For example, $\langle \overline{f} \rangle$ applies Eq.~(\ref{eq:bar}) but to the average quantity $\overline{f}$ on the larger averaging window $x \in [1.1,1.3]$. This larger averaging region should reduce sampling noise, at the cost of now averaging over a window in which (ensemble-averaged) quantities may change with position. By averaging first locally, dispersions are referenced to the local mean rather than the mean over a wider profile. For most quantities presented here, simply computing averages or dispersions in one step over the whole of the analysis region yields nearly identical results; there are noticeable but very modest differences for Figs. \ref{fig:fig_rho}, \ref{fig:fig_Anis}.

As a final note, where there is no reason for confusion, we use the overline for other averaging as well, as in the average pre-shock density $\overline{\rho}_0$, or in the average shock speed $\overline{U}_s$, where the average results from the linear fitting procedure for the time evolution of the average shock position, described in Sec. \ref{sec:setup}. 

\section{Results}\label{sec:results}

\begin{figure}
\includegraphics[width=\columnwidth]{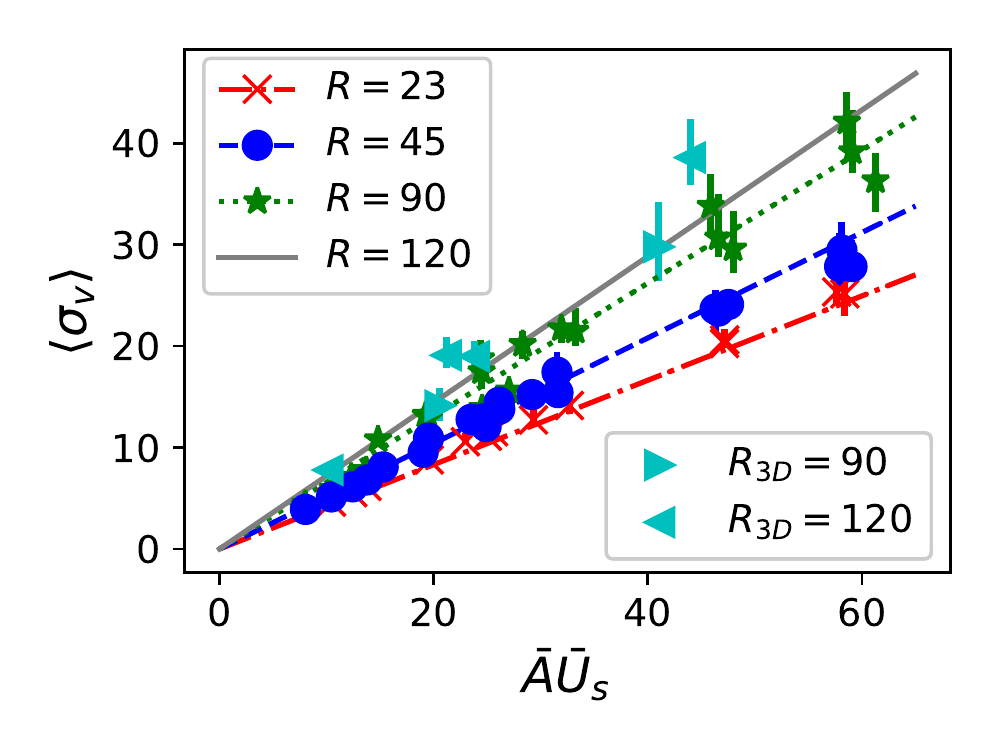}
\caption{\label{fig:fig_dv} Post-shock turbulent velocity dispersion (km/s) versus the product of pre-shock average Atwood number and mean shock speed (km/s). Each point is the result of analyzing a single simulation. Shown are a suite of 2D runs covering three different characteristic scales ($R = 23,\,45,\,90$) with many different values of the product $\overline{A} \, \overline{U}_s$. A smaller set of three-dimensional runs with $R = 90,\,120$ are also shown. The lines show, for each of the four different radii ($R = 23, 45, 90, 120$), the prediction of Eq.~(\ref{eq:deltav}). This relation is observed to reasonably predict the turbulent velocity dispersion, although with increasing spread as $R$ increases due to sampling effects (see text for further discussion and discussion of the 3D runs). The lines (error bars) associated with points in this and following figures show the spread in results obtained on moving the averaging window within the near post-shock region, see Sec. \ref{sec:averaging}.}
\end{figure}

In the context of young supernova remnants, prior work\citep{inoue2013} on the interaction of strong shocks with density non-uniformity suggests that the post-shock velocity dispersion is reasonably predicted by the result of linear analysis of the Richtmyer-Meshkov instability\citep{richtmyer1960}. Accordingly, with the average shock velocity $\overline{U}_s$, and defining an average Atwood number for the pre-shock material as $\overline{A} = (\sigma_\rho/\overline{\rho})_0/[1 + (\sigma_\rho/\overline{\rho})_0]$, the post-shock velocity dispersion, $\sigma_v \propto \overline{A} \, \overline{U}_s$.

Figure \ref{fig:fig_dv} shows the analysis-region velocity dispersion, $\langle \sigma_v \rangle$, versus the quantity $\overline{A} \, \overline{U}_s$ for our suite of 2D and 3D simulations. It is apparent that the proportionality predicted by the simple linear RMI result holds for the observed velocity dispersions. It is also apparent that the characteristic ``void" radius, $R$, of the pre-shock density non-uniformity has an influence on the turbulent velocity dispersion (we find that the shock speed itself, for a given drive, is relatively insensitive to $R$, consistent with \citet{kim2021}). 

Consider two cases that are closely matched in all quantities ($\overline{A}$, $\overline{U}_s$, $\overline{\rho}_0$, etc.) but that differ in $R$. When a flow's dynamics are governed by some velocity $v$ and lengthscale $l$ (viscosity is negligible), we have by dimensional analysis that the specific rate of energy change can be written $d E/d t = v^3/l$ (e.g., for turbulence forced in equilibrium at a lengthscale $l_f$ with an rms velocity $u_{rms}$, the rate of energy injection, or dissipation $\epsilon$, $d E/d t = \epsilon \propto u_{rms}^3/l_f$\citep{landau1987}). Suppose it were the case that the energy injection rate into the post-shock turbulence in these matched cases is comparable. At present, the maxima of the pre-shock density spectra are correlated with the void diameter, with this (nonuniform) density leading to the turbulence. The (large-scale) post-shock velocity is characterized by $\sigma_v$.
Then for matched cases analyzed at a fixed distance behind the average shock position, as is done here, we hypothesize $\sigma_{v} \propto R^{1/3}$. The trend lines shown in Fig. \ref{fig:fig_dv} for the 2D runs are,
\begin{equation}
\langle \sigma_{v,2D} \rangle = 0.52 (R/45)^{1/3} \overline{A} \, \overline{U}_s, \label{eq:deltav}
\end{equation}
and we observe the anticipated scaling with radius. Panel (c) of Figure \ref{fig:fig_3D} supports the notion that there is effectively forcing at $l_f \propto R$. While this dimensional analysis gives a scaling with radius consistent with the present simulation results, it is possible there is some alternate explanation which would also lead to consistent results. Note that, among other assumptions, we have assumed negligible viscosity; in the turbulence case this is only strictly true in the high Reynolds number limit. We should expect limitations on this scaling in radius, which we discuss further in Sec. \ref{sec:discussion}.

Note that the coefficient of proportionality in Eq.~(\ref{eq:deltav}), $0.52$ (chosen to approximately fit the data), will in general depend on choices we have made in our setup. Most simply, the turbulent velocity dispersion generally decreases with distance behind the average shock front (see (b) in Fig. \ref{fig:fig_snapshot}), and as a result the inferred proportionality coefficient will depend on the downstream distance of the post-shock averaging region. The post-shock velocity dispersion could depend on properties of the pre-shock density non-uniformity beyond those considered in Eq.~(\ref{eq:deltav}), that is, beyond $\overline{A}$ and a scale factor $R$. Such effects could change the coefficient of proportionality as well. 

\begin{figure}
\includegraphics[width=\columnwidth]{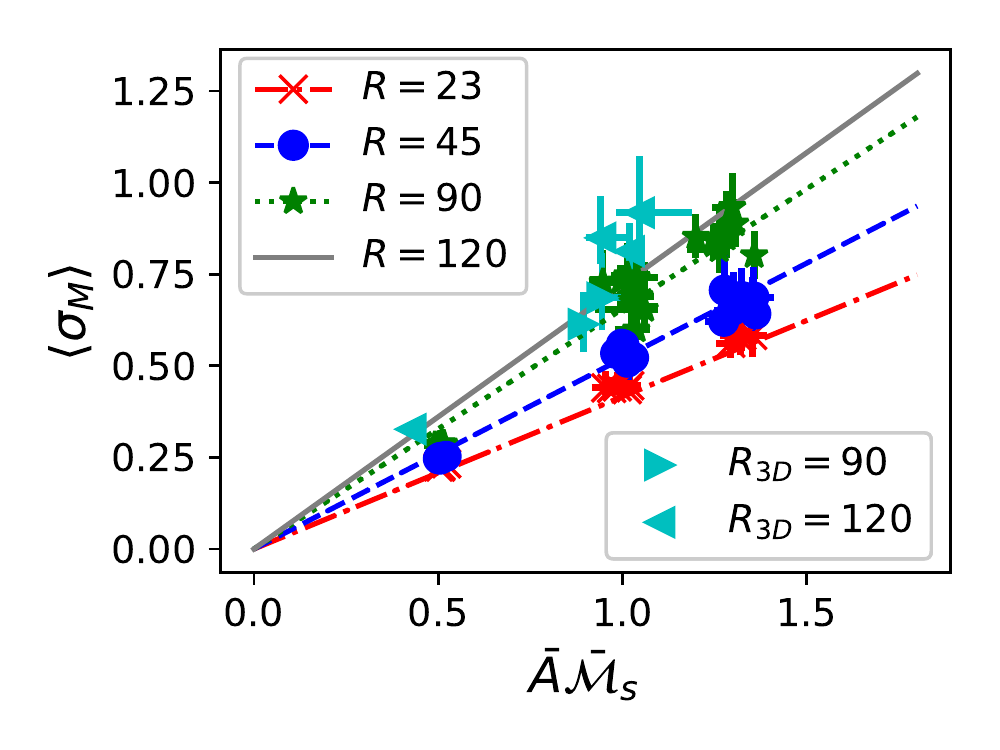}
\caption{\label{fig:fig_dM} A post-shock turbulent Mach number computed using the velocity dispersion and the average sound speed in the analysis region. The results closely follow the relation Eq.~(\ref{eq:M_t}), which is used to generate the trend lines in the figure (again with a slightly different coefficient for 3D runs). Compared with Fig. \ref{fig:fig_dv}, there is a clustering here of points with the same $\overline{A}$ but different $\overline{U}_s$ due to the fact that $\overline{\mathcal{M}}_s$ saturates while $\overline{U}_s$ does not, see Eq.~(\ref{eq:M_s}) and the surrounding discussion.}
\end{figure}

For example, although the quantity of data is too limited for a firm conclusion, the 3D cases in Fig. \ref{fig:fig_dv} are not very well fit by Eq.~(\ref{eq:deltav}) with a coefficient of $\sim0.52$, being better fit with a coefficient $\sim 0.6$ [and still following the overall scalings of Eq.~(\ref{eq:deltav})]. It would be reasonable to expect that this is a result of differences in the effective coefficient for shock interaction with spherical inhomogeneities in 3D compared with circular ones in 2D (that is, that the coefficient acts as a geometrical factor). However, as we discuss further shortly, we cannot conclude firmly at present that this is not simply a consequence of the (statistical) variation in the quantities in the averaging region (that is, after averaging multiple such 3D runs we may find this apparent discrepancy vanishes).

The number of voids contributing post-shock motion to the averaging region decreases as $R$ increases, since the domain size (and averaging window) is fixed for all cases. This decreasing sample is observed to lead to increasing scatter in $\langle \sigma_v \rangle$ as $R$ increases in the 2D cases, where we rerun some of the cases multiple times varying only the seed used to generate the random perturbations of the pre-shock void positions (see also Appendix \ref{sec:A}). In particular, in Fig. \ref{fig:fig_dv}, 2D runs with $\overline{A} \, \overline{U}_s \gtrsim 40$ are repeated multiple times with different randomized void shifts, with the $R= 90$ cases showing most spread around the trend line as expected.
As a final note on the proportionality coefficient, \citet{inoue2013}, working in 3D, find a coefficient of 1 for a density non-uniformity with a power spectrum following an isotropic $k^{-5/3}$ power law (maximum at the simulation box size); see also the discussion in Sec.~\ref{sec:discussion}. 

In accordance with the results shown in Fig. \ref{fig:fig_dv}, we may exert a measure of control on post-shock turbulent velocities in such a shock-tube setup by altering the pre-shock density structure (see also Appendix \ref{sec:A}) and the velocity of the shock. A second aspect of the post-shock turbulence we may wish to influence is the turbulent Mach number, which quantifies the compressibility of the turbulence. As we show shortly, there is a measure of independent control for the turbulent velocity and Mach number.

Dividing Eq.~(\ref{eq:deltav}) by the post-shock sound speed averaged over the analysis region, $\langle \overline{C}_s \rangle$, we will find a turbulent Mach number
\begin{equation}
\langle \sigma_M \rangle = 0.52 (R/45)^{1/3} \overline{A} \, \overline{\mathcal{M}}_s. \label{eq:M_t}
\end{equation}
From Eq.~(\ref{eq:M_t}), we see that $\langle \sigma_M \rangle = \langle \sigma_v \rangle/\langle \overline{C}_s \rangle$ can be influenced by the pre-shock density ($\overline{A}$) and is also proportional to the shock Mach number with respect to the \emph{post-shock} sound speed. We denote this Mach number $\overline{\mathcal{M}}_s = \overline{U}_s/\langle \overline{C}_s \rangle$. As for the shock velocity, the overbar here is to recall the fact that there is only an average shock front position and that the velocity of this front decays somewhat during propagation. 

\begin{figure}
\includegraphics[width=\columnwidth]{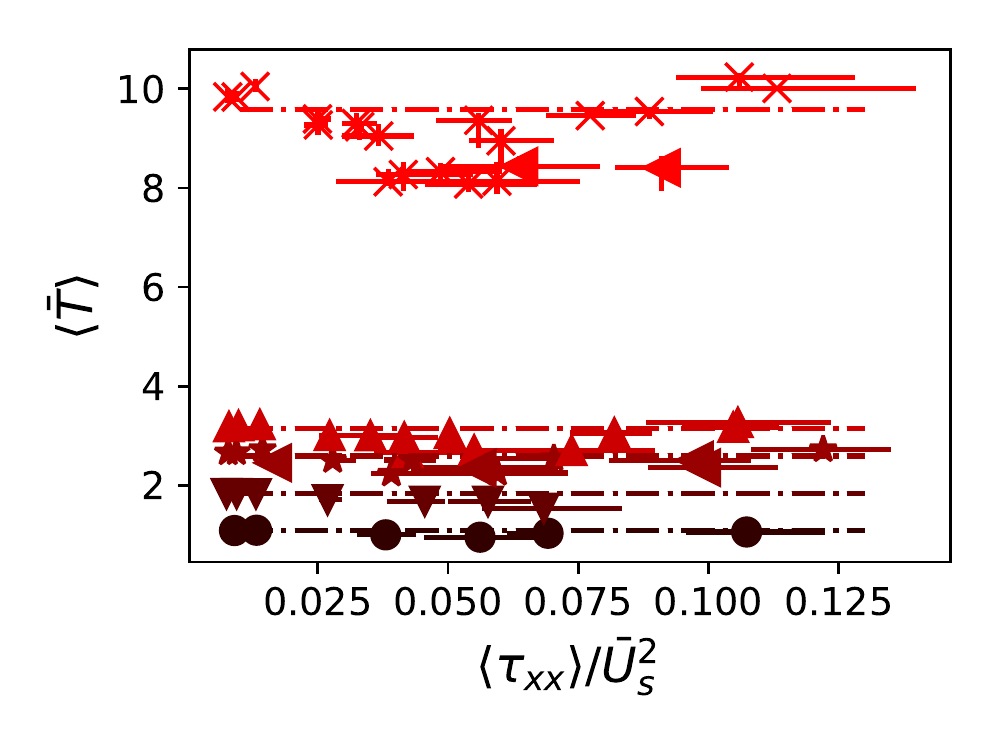}
\caption{\label{fig:fig_T} The post-shock temperature (eV), volume averaged over the analysis region, plotted against a measure of the (downstream) turbulent pressure relative to the shock ram pressure. The post-shock temperatures cluster based on the shock drive strength, $\overline{M}_s \sim 14$ (circle, black) , $\overline{M}_s \sim 16$ (down triangle), $\overline{M}_s \sim 20$ (star), $\overline{M}_s \sim 22$ (up triangle), $\overline{M}_s \sim 37$ (x, brightest red). Also shown are the 3D run results for $\overline{M}_s \sim 20$ (left triangle) and $\overline{M}_s \sim 37$ (left triangle, brightest red), which have the same behavior as the 2D runs. For each drive strength, a horizontal dashed line shows the post-shock temperature averaged over the analysis region when the pre-shock is uniform ($\overline{A} = 0 $). The post-shock volume-averaged temperature is observed to be insensitive to the relative (to the shock strength) turbulence intensity generated in the interaction.}
\end{figure}

Figure \ref{fig:fig_dM} shows $\langle \sigma_M \rangle$ versus $\overline{A} \, \overline{\mathcal{M}}_s$ for the suite of runs. We can observe that the relation Eq.~(\ref{eq:M_t}), which is used to generate the trend lines, fits the simulation suite reasonably well (with again some uncertainty for the 3D cases). Note that this characteristic turbulent Mach number, $\langle \sigma_M \rangle$, being defined in terms of an average rather than a local sound speed, will only in general equal the average of the local Mach number for the isothermal turbulent fluctuation case, which the present case is not. The probability distributions and correlations of density and sound speed in a similar setup are analyzed in detail in \citet{dhawalikar2021}. We also note that the present Mach number is a volume-weighted one, which in general can be different than the mass-weighted result.

In comparing Fig. \ref{fig:fig_dv} to Fig. \ref{fig:fig_dM}, we observe a clustering of certain points in the latter. This is a result of the saturation of $\overline{\mathcal{M}}_s$ in Eq.~(\ref{eq:M_t}), such that different values of $\overline{U}_s$ yield similar values of $\overline{\mathcal{M}}_s$. Then, variation in $\sigma_M$ is driven primarily by variation in $\overline{A}$. We now elaborate on this saturation of $\overline{\mathcal{M}}_s$.

In the most basic analysis, considering the case of an unperturbed 1D shock, we can use the Rankine-Hugoniot conditions\citep{landau1987} to write $\mathcal{M}_s$ as
\begin{equation}
\mathcal{M}_s = \frac{(\gamma + 1) M_s^2}{\sqrt{(2 \gamma M_s^2 - \gamma + 1)((\gamma - 1) M_s^2 + 2)}}. \label{eq:M_s}
\end{equation}
Here $M_s$ is the usual shock Mach number with respect to the pre-shock medium.

In the limit of strong shocks where the terms in $M_s^2$ dominate, we have from Eq.~(\ref{eq:M_s}) that $\mathcal{M}_s \rightarrow (\gamma + 1)/\sqrt{2 \gamma(\gamma-1)}$. That is, while the shock Mach number is unbounded, $\mathcal{M}_s$ saturates at a $\gamma$ dependent value; for $\gamma = 5/3$ this value is $4 /\sqrt{5} \approx 1.79$. 

Consider then the ratio $\sigma_v/\sigma_M \sim \overline{U}_s/\overline{\mathcal{M}}_s$. Since $\mathcal{M}_s$ will be near its saturation value for a wide range of achievable shock Mach numbers (shock speeds), the resulting turbulence can have similar $\sigma_M$ values at substantially different turbulent velocity dispersions $\sigma_v$. Viewed in a different light, this means that, so far as increasing this turbulent Mach number is concerned in the present regime (as controlled by the simulated equations), there is a rapidly diminishing return to driving with higher-Mach shocks as $\mathcal{M}_s$ saturates in Eq.~(\ref{eq:M_s}). Note that as $\gamma \rightarrow 1$ this constraint is relaxed.

\begin{figure}
\includegraphics[width=\columnwidth]{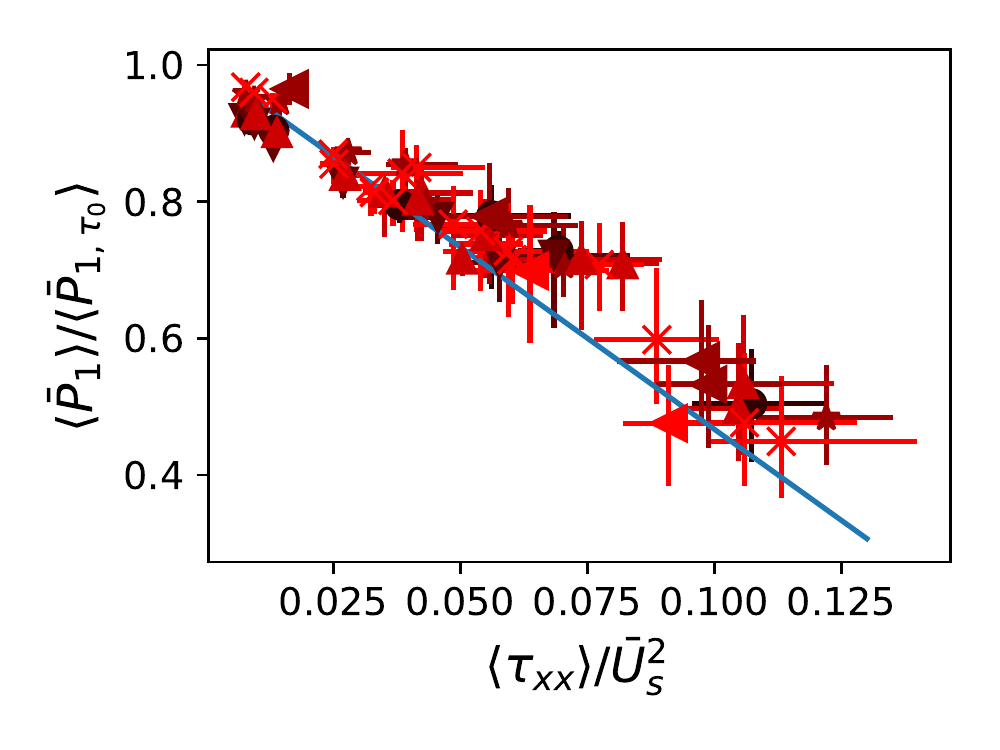}
\caption{\label{fig:fig_P} As Fig. \ref{fig:fig_T}, but showing the analysis-region-averaged pressure rather than the temperature. Here for each drive case we normalize the pressure by its value in the unperturbed (uniform upstream density) case, which collapses the different drive cases. The solid (blue) line shows the prediction of Eq.~(\ref{eq:p_jump_2}) with $d=3$ and $\alpha = 1$.}
\end{figure}

In utilizing the unperturbed 1D shock relations, the preceding analysis ignored any possible corrections to this post-shock Mach number arising from the shock becoming disrupted by the density non-uniformity. That is, if the turbulence generation came at the expense (in part) of post-shock temperature, then one might expect a reduction in the post-shock sound speed and a corresponding increase in $\overline{\mathcal{M}}_s$ over the 1D saturation value.

In the case of the present volume-averaged Mach number, no such feedback is discernible with the current suite of simulations. Figure \ref{fig:fig_T} shows the post-shock temperature averaged over the analysis region, again using volume weighting. Here the average temperature is plotted for each case against the quantity $\langle \tau_{xx} \rangle/\overline{U}_s^2$, which is the ratio of the $xx$ component of the density-weighted Reynolds stress tensor to the square of the shock speed. This quantity represents the relative contributions of the turbulence and shock to a momentum conservation equation across the average shock front; we discuss such an analysis shortly. Figure \ref{fig:fig_T} shows that the volume-averaged post-shock temperature appears insensitive to the relative amount of turbulence ($\langle \tau_{xx} \rangle/\overline{U}_s^2$).

Here we have focused on the behavior of the volume-averaged sound speed and the associated turbulence and post-shock Mach numbers. Below, we analyze the volume-averaged post-shock pressure and density, which are related to the density-weighted temperature through the (ensemble) averaged ideal gas law, $\overline{p} = \overline{\rho} R \tilde{T}$. Here $R$ is the ideal gas constant as usual. As we will find, these quantities are impacted by the degree of disruption of the shock (as measured by $\langle \tau_{xx} \rangle/\overline{U}_s^2$). As a result, we should expect that, if the density-weighted post-shock Mach number is of interest, the limit imposed by Eq.~(\ref{eq:M_s}) can be relaxed by the reduction in post-shock $\tilde{T}$ for increasing ``void" radius and $\overline{A}$.

\begin{figure}
\includegraphics[width=\columnwidth]{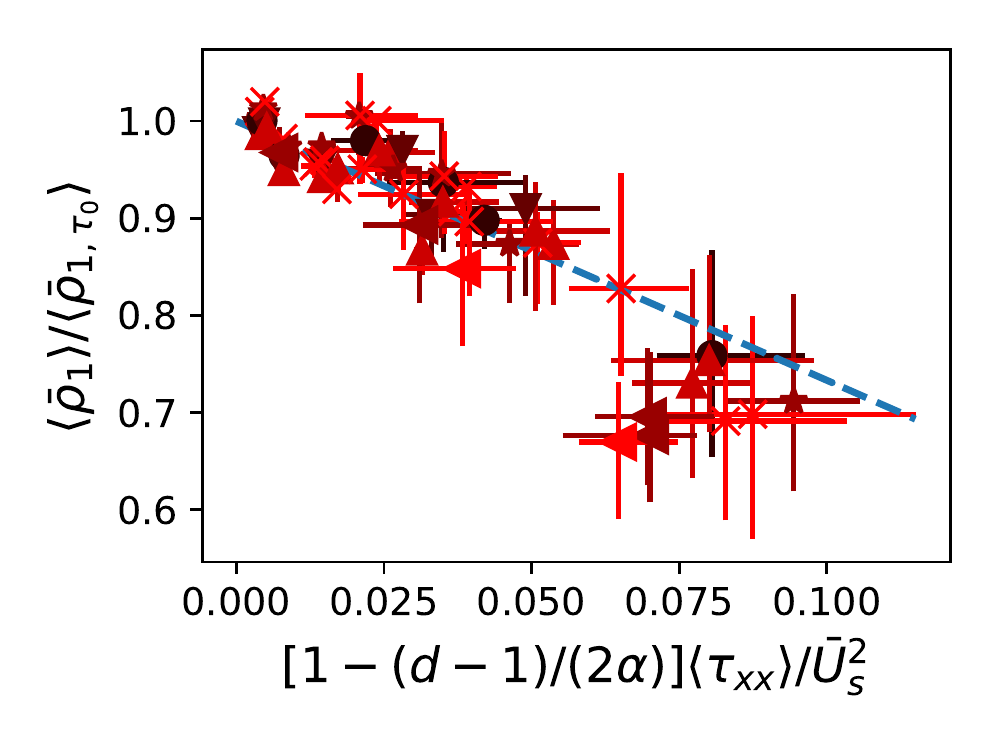}
\caption{\label{fig:fig_rho} As Fig. \ref{fig:fig_P} and Fig. \ref{fig:fig_T}, but showing the analysis-region-averaged density jump across the shock relative to the density jump in the unperturbed case (when averaged over the analysis region, the unperturbed cases yield density jumps $\sim 3.3$ - $3.8$). Here the subscript $\tau_0$ indicates the unperturbed case, and the subscript $1$ is post shock. The density jump across the shock is more sensitive to the anisotropy $\alpha$ than is the pressure jump, and so here we plot against the quantity $[1 - (d-1)/(2 \alpha)] \langle \tau_{xx} \rangle/\overline{U}_s^2$ which arises in Eq.~(\ref{eq:rho_jump_2}). The (blue) dashed line shows Eq.~(\ref{eq:rho_jump_2}). For brevity, we label the $y$ axis as the ratio of the (analysis-region averaged) post-shock density in the perturbed case versus the unperturbed case, $y = \langle \rho_{1} \rangle/\langle \rho_{1,\tau_0} \rangle$. However, we actually compute the ratio of density jumps between the two cases, $y = (\langle \rho_{1} \rangle/\overline{\rho}_0)/(\langle \rho_{1,\tau_0} \rangle)/\rho_{0,\tau_0})$, which can differ somewhat because, while the unperturbed pre-shock density $\rho_{0,\tau_0} = 0.05$ g/cm$^3$ everywhere, the average pre-shock density in the perturbed cases is only approximately $0.05$, $\overline{\rho}_0 \approx 0.05$ g/cm$^3$.}
\end{figure}

One may wonder how well the post-shock quantities apart from the turbulent velocity, which we have already discussed, can be predicted with a simplified treatment that takes the present extended shock transition to be a single (average) normal shock. Following work on shock-turbulence interaction\citep{lele1992,larsson2013}, we write the (Favre) averaged conservation equations across such a normal shock,
\begin{align}
\llbracket \overline{\rho} \tilde{v}_x \rrbracket &= 0, \label{eq:con1} \\
\llbracket \overline{\rho} \tilde{v}_x^2 + \overline{p} + \overline{\rho} \tau_{xx} \rrbracket &= 0, \label{eq:con2} \\
\llbracket \frac{\overline{\rho} \tilde{v}_x^3}{2} + \frac{\gamma}{\gamma - 1} \overline{p}\tilde{v}_x + \overline{\rho} \tilde{v}_x \left( \tau_{xx} + \frac{\tau_{ii}}{2}\right) \rrbracket &\approx 0. \label{eq:con3} 
\end{align}
In Eqs.~(\ref{eq:con1})-(\ref{eq:con3}), $\llbracket \rrbracket$ denotes jumps in the quantity across the shock front, for example, $\llbracket f \rrbracket = f_{1} - f_{0}$, with the subscripts denoting pre (0) and post (1) shock. In writing Eq.~(\ref{eq:con3}) we have dropped a turbulent triple correlation term and the temperature-velocity correlation. At present, we assume axisymmetry (for 3D), $\tau_{yy} = \tau_{zz}$, and express the trace of the Favre-averaged Reynolds stress, $\tau_{ii}$, in terms of an anisotropy $\alpha = \tau_{xx}/\tau_{yy}$. Then, $\tau_{ii} = \tau_{xx} \left(1 + (d - 1)/\alpha \right)$, where $d=2$ for 2D and $d=3$ for 3D.

In the present case there is no turbulence pre-shock. To simplify the calculation, we drop the pre-shock pressure, which is small for all our cases compared to the ram pressure associated with the shock (we work in a frame where the average shock is stationary). Taking the post-shock turbulence $\tau_{xx}$ and anisotropy $\alpha$ as given, Eqs.~(\ref{eq:con1})-(\ref{eq:con3}) can be solved to give the post-shock (average) density, pressure, and flow velocity. At present, we power expand these solutions in the quantity $\tau_{xx}/\overline{U}_s^2$, which is observed from Fig.~\ref{fig:fig_T} to be (reasonably) small for all our cases.

Using the power-expanded solutions, we normalize the post-shock quantities to their values in the unperturbed case ($\tau_{ij} = 0$, denoted $\tau_0$). To first order, the (normalized) post-shock density and pressure are,
\begin{align}
\frac{\overline{\rho}_1}{\overline{\rho}_{1,\tau_0}} &\approx  1 - \frac{2 \alpha (3 - \gamma) - 2(d - 1)(\gamma - 1)}{4 \alpha}\frac{\gamma + 1}{\gamma - 1} \frac{\tau_{xx}}{U_s^2}, \label{eq:rho_jump} \\
\frac{\overline{p}_1}{\overline{p}_{1,\tau_0}} &\approx 1 + \frac{(d - 1) (\gamma - 1)^2 + \alpha (1 - 6\gamma + \gamma^2)}{4 \alpha}\frac{\gamma + 1}{\gamma - 1}\frac{\tau_{xx}}{U_s^2} . \label{eq:p_jump} 
\end{align}
When $\gamma = 5/3$, as for our suite of simulations, one finds from Eqs.~(\ref{eq:rho_jump}), (\ref{eq:p_jump}),
\begin{align}
\frac{\overline{\rho}_1}{\overline{\rho}_{1,\tau_0}} &\approx 1 - \frac{8}{3} \left(1 - \frac{d-1}{2 \alpha} \right) \frac{\tau_{xx}}{U_s^2} , \label{eq:rho_jump_2} \\
\frac{\overline{p}_1}{\overline{p}_{1,\tau_0}} &\approx 1 - \frac{8}{9}\left( 7 - \frac{d - 1}{2 \alpha} \right) \frac{\tau_{xx}}{U_s^2}. \label{eq:p_jump_2} 
\end{align}

Figures \ref{fig:fig_P} and \ref{fig:fig_rho} use the simulations to investigate the post-shock pressure and density, respectively, relative to the unperturbed cases. In each case, the predictions of Eqs.~(\ref{eq:rho_jump_2}), (\ref{eq:p_jump_2}) are also plotted. Overall, there is fair agreement between these theoretical results using a single average normal shock and the simulation results. We now discuss some features of the pressure and density results.

\begin{figure}
\includegraphics[width=\columnwidth]{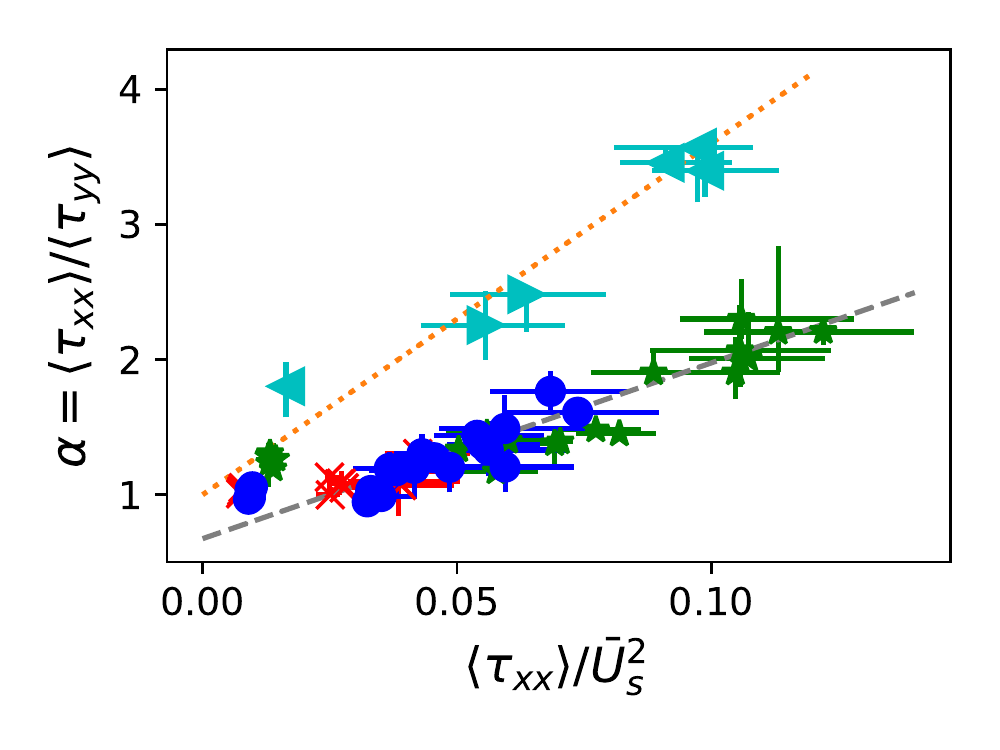}
\caption{\label{fig:fig_Anis} A measure of the turbulence anisotropy $\alpha$ in the post-shock analysis region versus $\langle \tau_{xx} \rangle/\overline{U}_s^2$. Case symbols follow the legend in Fig. \ref{fig:fig_dv}. Two trend lines are plotted, one (dashed, gray) for the 2D cases and one (dotted, orange) for the 3D cases. The slopes differ by a factor of two, see the discussion near the end of Sec. \ref{sec:results}.}
\end{figure}

The pressure result, Eq.~(\ref{eq:p_jump_2}), suggests that the pressure jump is relatively insensitive to the anisotropy (for $\alpha \gtrsim 1$) and dimensionality, especially compared to the density result in Eq.~(\ref{eq:rho_jump_2}). We observe this in the simulation results in Fig. \ref{fig:fig_P}, which are relatively collapsed to a line even though we ignore the influence of $d$ and $\alpha$ when plotting them and we also plot Eq.~(\ref{eq:p_jump_2}) for a single pair of $d$, $\alpha$. The density result is more sensitive to $d$ and $\alpha$, and we include these factors as they enter in Eq.~(\ref{eq:rho_jump_2}) when plotting in Fig. \ref{fig:fig_rho}.

Figure \ref{fig:fig_P} shows that the post-shock pressure is more than halved from the unperturbed case in our simulations with the largest turbulent contribution relative to shock ram pressure (largest $\langle \tau_{xx} \rangle/\overline{U}_s^2$). As previously noted, the density-weighted average temperature, $\tilde{T}$, is related to the volume averaged pressure and density through the ideal gas equation of state, $\overline{p} = \overline{\rho} R \tilde{T}$. It then depends on $\tau_{xx}/\overline{U}_s^2$.

We turn now to the anisotropy of the turbulence generated in this shock interaction with non-uniform density. As discussed above, the basic theory represented by Eqs.~(\ref{eq:rho_jump}), (\ref{eq:p_jump}) predicts that the density and pressure jumps are more or less sensitive to the anisotropy. We note that the relation Eq.~(\ref{eq:deltav}) also fits the mass-weighted velocity dispersion $\langle \sqrt{\tau_{ii}} \rangle$ reasonably well for the current suite of simulations. Then, this result, if combined with the anisotropy $\alpha$, can be used to ``close" the shock jump relations Eqs.~(\ref{eq:con1})-(\ref{eq:con3}), with the understanding that we have ignored the triple velocity correlation and the temperature-velocity correlation.

At present we simply construct trend lines for the anisotropy in the suite of simulations. Figure \ref{fig:fig_Anis} shows the averaged anisotropy $\alpha = \langle \tau_{xx} \rangle/ \langle \tau_{yy} \rangle$ plotted against the relative turbulent momentum contribution $\langle \tau_{xx} \rangle/\overline{U}_s^2$. We observe that the anisotropy generally increases with $\langle \tau_{xx} \rangle/\overline{U}_s^2$ (although at low values there is a saturation at $\alpha \sim 1$ in the 2D cases).

The trend line shown in Fig. \ref{fig:fig_Anis} for the 2D cases is $\alpha \sim 1 + 13 (\langle \tau_{xx} \rangle/\overline{U}_s^2 - 1/40)$. For the 3D cases, the trend line shown is $\alpha \sim 1 + 26 \langle \tau_{xx} \rangle/\overline{U}_s^2$, so that the 3D case trend line has double the slope. This is expected on the grounds of dimensionality. Recalling $\tau_{ii} = \tau_{xx} \left(1 + (d - 1)/\alpha \right)$, suppose we consider matched 2D and 3D cases. Since Eq.~(\ref{eq:deltav}) fits the simulation suite's mass-weighted velocity dispersions reasonably well, and the 2D and 3D cases are not clearly distinguishable in this regard, take $\tau_{ii,2D} \sim \tau_{ii,3D}$. Then we have $\alpha_{3D} \sim 2 \alpha_{2D}$.

Note that as the analysis region is moved further downstream (to the right in Fig. \ref{fig:fig_snapshot}), we observe that the anisotropy tends to decrease towards equilibrium (not shown). A similar behavior is observed in wind tunnel shock-turbulence experiments\citep{barre1996}. Since the different ``void" radius cases have different characteristic turbulence scales, the rate of change of this and other quantities with downstream distance may differ from one radius case to another.

\section{Discussion} \label{sec:discussion}

Although we have shown here a degree of control over turbulence generation through varying the average Atwood number $\overline{A}$ of the pre-shock state, there may be significant freedom yet to explore in tailoring the pre-shock density state to generate a desired turbulent state. Now we briefly outline a few ways in which this could be useful, as motivation for future work. 

At present the pre-shock density perturbations are introduced in a spatially homogeneous fashion. One could, instead, vary $\overline{A}$ (or the void radius $R$) spatially, with the goal of introducing or eliminating spatial gradients in the turbulence intensity (or turbulence lengthscale). For example, Fig.~\ref{fig:fig_snapshot} shows that the turbulent velocity dispersion $\sigma_v$ decreases with distance behind the (average) shock. Could one reduce this effect by beginning with $\overline{A}(x)$? In a similar fashion, one might study turbulence with gradients in turbulence intensity (or gradient in lengthscale) by beginning with $\overline{A}$ varying in one or more of the transverse directions, $\overline{A}(y)$ (or $R(y)$).

Note that to generate our higher-density-variance cases ($(\sigma_\rho/\overline{\rho})_0 \sim 1.9$ for 2D cases), we have introduced, in addition to the ``voids" seen in Fig. \ref{fig:fig_snapshot}, smaller radius ``beads" of higher density than the initially uniform background (see Fig. \ref{fig:fig_den} (b) in Appendix \ref{sec:B}). Strictly speaking this introduces a second (and smaller) radius into the problem, which may somewhat increase the scatter observed in various figures with respect to the trend lines, but evidently does not overwhelm the main scaling in $R$ in Figs. \ref{fig:fig_dv}, \ref{fig:fig_dM}. Nonetheless, it may be worthwhile in the future to consider alternate methods to introduce any such third density into the problem when trying to raise $\overline{A}$. It should be kept in mind that the introduction of multiple materials with different equations of state into the pre-shock medium may alter at least certain present results.

The shock-tube setup studied here appears well suited for generating a specified volume of turbulent flow. One could also consider, instead, more spatially inhomogeneous setups. In the neutral gas case, for example, there is the turbulence generated in jet mixing layers\citep{belan2010}. The generation of turbulence by jets is also of interest in the astrophysical context, for example through instability in heliospheric jets interacting with the interstellar medium\citep{opher2021}.

Another area of future interest is the generation of post-shock magnetohydrodynamic (MHD) turbulence for a shock propagating through a medium with non-uniform density. Let us make a simple estimate of the maximum possible average magnetic fields in the present cases by assuming the post-shock turbulent energy is instead partitioned equally into turbulent energy and magnetic energy (that is, we assume a magnetic energy density of half the original turbulent energy density). Given post-shock mean densities on the order of $\gtrsim 0.1$ g/cm$^3$, and fluctuating velocities $\lesssim 40$ km/s, we find in this simple estimate magnetic fields $\sim10-100$ T.

The ratio of post-shock thermal pressure to post-shock turbulent pressure (``turbulent $\beta$'') for the present suite of simulations varies from $\sim1 - 25$. As such, continuing with the above simple estimate, we would find magnetic $\beta$ (the ratio of the thermal pressure to magnetic pressure) of $\gtrsim 2$; this result is similar to the results of ideal MHD calculations of a shocked non-uniform medium in a different context\citep{inoue2011}. Note that we do not claim such fields would necessarily be present for the cases studied in this work. In the present cases the turbulent $\beta$ tends to be lower (the inferred relative magnetic field is higher) for lower Mach number shocks where the post-shock is cooler. We should expect to need to consider the conductivity of such plasmas (which is important also in HED shock tube designs looking at magnetized Rayleigh-Taylor instability\citep{barbeau2021}), as well as whether there are sufficient seeds for the magnetic field (or the use of a pre-imposed initial magnetic field). 

In the present simulations, where the material in which the turbulence is generated is treated as an ideal gas with $\gamma = 5/3$, our results suggest a limit to the achievable volume-weighted Mach numbers. From Eq.~(\ref{eq:M_t}), we have that $\sigma_M$ is proportional to $\overline{A}$, $\overline{\mathcal{M}}_s$, and a combined coefficient that, at a minimum, depends on properties of the pre-shock density beyond $\sigma_\rho/\overline{\rho}$. As we have already discussed, when $\gamma = 5/3$, $\mathcal{M}_s$ saturates in the unperturbed 1D theory at $\sim1.79$, and it appears to also saturate in the perturbed (non-uniform pre-shock density) case. This saturation however can be relaxed if one is concerned with density-weighted quantities.

By definition $\overline{A} = (\sigma_\rho/\overline{\rho})_0/[1 + (\sigma_\rho/\overline{\rho})_0]$ also has a maximum achievable value of 1. Although it is not apparent as written, we should expect the combined coefficient, $0.52 (R/45)^{1/3}$, to similarly have a maximum achievable value as a consequence of (average) momentum and energy conservation. In other words, the total possible turbulence generation will be limited by the available shock energy and momentum. In practical terms, the scaling in $R$ must either saturate at or before the domain size; for the ``voided" pre-shock density construction here, we aim for many voids across the domain for statistical purposes (with too few voids we may also question whether the resulting state will be ``turbulent"). As previously noted, \citet{inoue2013} finds a proportionality coefficient of 1 ($\sigma_v = \overline{A} \, \overline{U}_s$) using a power-law pre-shock density spectrum peaked at the simulation box size.

One apparent way to make it easier to achieve higher Mach numbers is to reduce $\gamma$. In the context of HED plasmas, $\gamma$ could be reduced by operating in a regime with substantial radiative loss in the post-shock plasma. There is the possibility in the current setup of doping the ``working" material with the aim of increasing the amount of radiative loss. 

Since a reduction in $\gamma$ increases the compressibility, we should expect the extent of the post-shock region to be reduced, which may make the analysis more difficult. Indeed, simulations of the present setup using a (tabular) carbon equation of state for the foam show compression ratios in the unperturbed cases that well exceed the maximum of $\rho_{1,\tau_0}/\rho_{0,\tau_0} \rightarrow 4$ for the case with $\gamma = 5/3$. One may anticipate a reduced effective $\gamma$, leading to higher compression, for such cases where the material can ionize substantially across the shock. 

Regarding Eq.~(\ref{eq:deltav}), which states that $\sigma_v \propto R^{1/3} \overline{A} \, \overline{U}_s$, we also comment that the original theoretical result for the linear RMI fits better when the \emph{post}-shock Atwood number is utilized rather than the pre-shock Atwood number\citep{richtmyer1960} (see also the discussion in \citet{brouillette2002}). Here we consider pre-shock quantities for ease of developing predictive capabilities (and because here the post-shock average Atwood number should be determined self-consistently with the turbulence). However, even without using directly a post-shock Atwood number, it is possible that an alternate average Atwood number could be constructed that would lead to improved predictive power for $\sigma_v$ once one considers more general cases of turbulence generation by shock interaction. Such possible cases include weaker shocks or a pre-shock medium that contains more than a single material (is governed by a different equation of state or $\gamma$ in different regions). In these cases it might prove useful to make some accounting in the average Atwood number for the expected (local) density jump varying for different pre-shock regions (even ignoring turbulence).

\section{Summary} \label{sec:summary}
Using numerical simulations we have studied the turbulence generated by a strong shock passing through a medium with highly non-uniform density. The medium and the drive of the shock are constructed to be in the spirit of those that might be achieved in an HED shock-tube experiment. We find that the post-shock turbulent velocity and volume-weighted turbulent Mach number (using an average sound speed) are reasonably predicted by combining theoretical considerations with fits for the present simulations. With the turbulence known, we find that an analysis based on a single normal shock also reasonably predicts the post-shock averaged pressure and density. Since the turbulence production in the simulations responds in a reasonably predictable way to changes in the pre-shock density structure and the shock-drive, both of which are controllable, we may consider the present setup as the basis for a scheme for controlled turbulence generation in the laboratory. Planned future work will utilize the present results to aid in the design of an HED shock-tube platform for turbulence generation.

\acknowledgements
This work was performed under the auspices of the U.S. Department of Energy by the Lawrence Livermore National Laboratory under Contract No. DE-AC52-07NA27344. S.D. was supported by the LLNL-LDRD Program under Project No. 20-ERD-058.

C.F.~acknowledges funding provided by the Australian Research Council (Future Fellowship FT180100495), and the Australia-Germany Joint Research Cooperation Scheme (UA-DAAD). C.F.~further acknowledges high-performance computing resources provided by the Australian National Computational Infrastructure (grant~ek9) in the framework of the National Computational Merit Allocation Scheme and the ANU Merit Allocation Scheme, and by the Leibniz Rechenzentrum and the Gauss Centre for Supercomputing (grant~pr32lo).

The authors would like to thank Bruce A. Remington, Mario J.-E. Manuel and Otto L. Landen for supporting the project, and would also like to thank N. J. Fisch for raising the possibility of varying $\overline{A}$ as a function of space in pre-shock density profiles.

This document was prepared as an account of work sponsored by an agency of the United States government. Neither the United States government nor Lawrence Livermore National Security, LLC, nor any of their employees makes any warranty, expressed or implied, or assumes any legal liability or responsibility for the accuracy, completeness, or usefulness of any information, apparatus, product, or process disclosed, or represents that its use would not infringe privately owned rights. Reference herein to any specific commercial product, process, or service by trade name, trademark, manufacturer, or otherwise does not necessarily constitute or imply its endorsement, recommendation, or favoring by the United States government or Lawrence Livermore National Security, LLC. The views and opinions of authors expressed herein do not necessarily state or reflect those of the United States government or Lawrence Livermore National Security, LLC, and shall not be used for advertising or product endorsement purposes.

\appendix
\section{Pre-shock density}\label{sec:A}
Here we provide some additional detail on the non-uniform density that exists prior to the shock propagation. The overarching goal is to support the main physics analysis by achieving different values of the (initial) density variance $(\sigma_\rho/\overline{\rho})_0$ with approximately fixed averaged density ($\overline{\rho}_0$). At the same time, although it is not the focus of the present work, we choose constructions that are in the spirit of what might be possible for HED shock-tube targets through existing techniques\citep{hamilton2016,saha2018}. 

\begin{figure}
\includegraphics[width=\columnwidth]{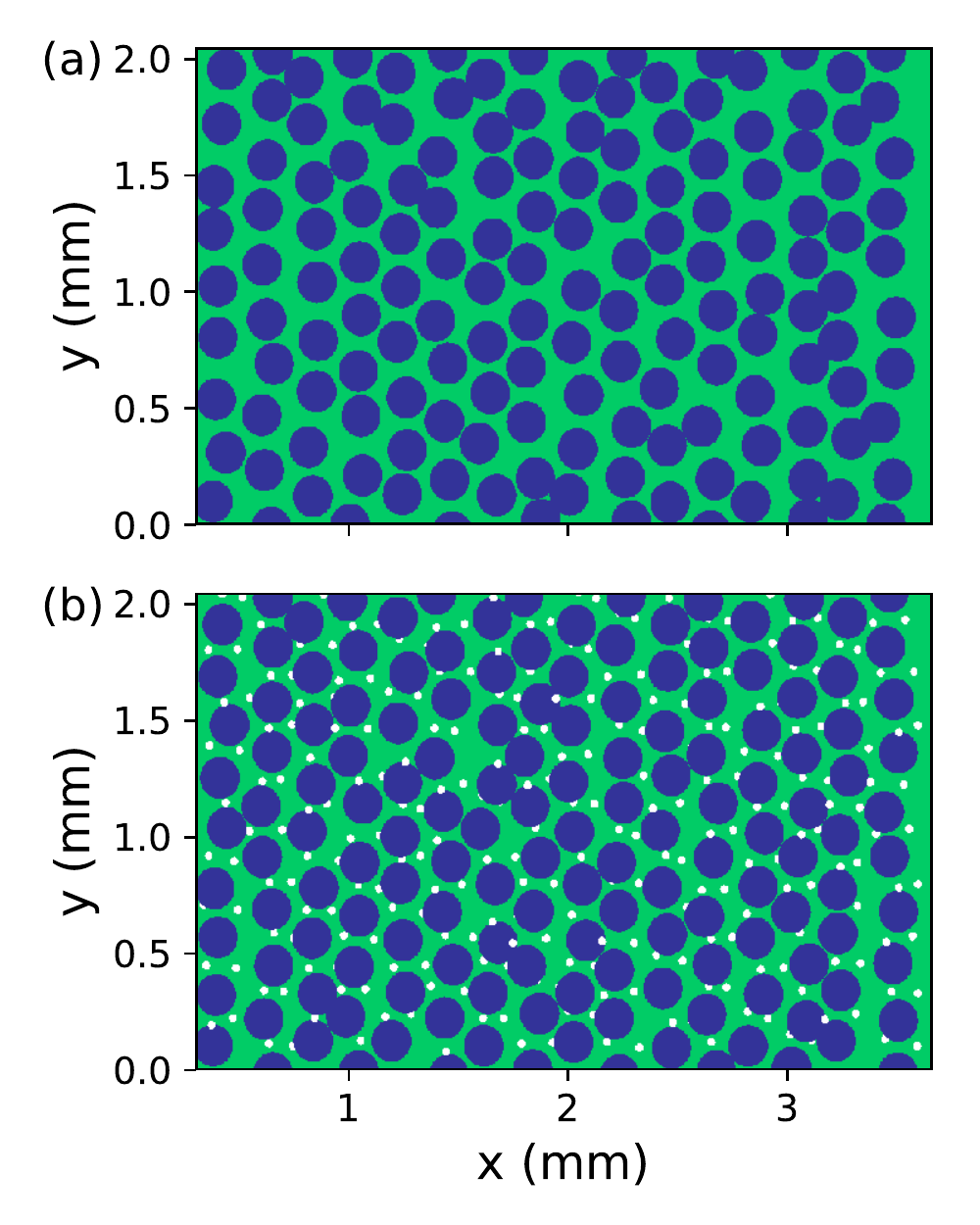}
\caption{\label{fig:fig_den} Examples of initial density field constructions for the different density variance cases in 2D. See text of Appendix \ref{sec:A}.}
\end{figure}

As noted in Sec.~\ref{sec:approach}, we create three different conditions for the normalized density variance $(\sigma_\rho/\overline{\rho})_0 \sim 0.38, 1.05, 1.9$ (for the 2D runs),  while $\overline{\rho}_0 \sim 0.05$  g/cm$^3$ in all cases. For the cases with density variance of about 1, the pre-shock density is created by ``voiding" an initially uniform density ($\rho = 0.1$  g/cm$^3$) such that roughly half the area is taken up by circular ``voids". In the simulations, we set the density in these ``void" regions to be a very small value ($10^{-3}$  g/cm$^3$) rather than zero. Panel (a) of Fig. \ref{fig:fig_den} shows a 2D example of such a case. Here we arrange the centers of the 2D spheres logically following a single layer of a hexagonal-close-packed (HCP) lattice, but with the spacings relaxed to achieve the targeted 50\% void fraction. The centers are then randomly displaced in each Cartesian direction by a value drawn from a uniform distribution with a maximum (plus and minus) similar to the unperturbed inter-void spacing along that direction. Note that we orient the HCP lattice such that there are no lines of fixed $y$ (or fixed $y$ and $z$ in 3D) that pass through the region without intersecting both voids and material.

To create the lower initial density variance cases, $(\sigma_\rho/\overline{\rho})_0 \sim 0.38$, we use the same process just described, but with a lower initial uniform density ($\rho = 0.066$  g/cm$^3$) and a higher ``void" density, such that the ``voids" are instead simply a somewhat lower density ($\rho = 0.03$  g/cm$^3$). 

The higher density variance cases $(\sigma_\rho/\overline{\rho})_0 \sim 1.9$ also follow a similar procedure as the $(\sigma_\rho/\overline{\rho})_0 \sim 1.05$ cases, but additionally make use of small regions of higher density (``beads") to increase the density variance. These regions are circular (2D) or spherical (3D), and have a density of 4x the initial uniform density (so $\rho = 0.4$  g/cm$^3$). In the 2D cases, an example of which appears in panel (b) of Fig. \ref{fig:fig_den}, there are two such beads (small, white, circles) for each void, with a radius of $R/5$. These beads are positioned, with random perturbations, in the region between the voids.

The 3D cases are in general similar to the 2D ones, with spherical ``voids" and ``beads" substituting for circular ones and the ``void" arrangement following a relaxed and randomly perturbed HCP lattice. Most spherical cases use a somewhat lower void volume fraction, such that, with empty (very low density) voids and $\rho = 0.1$ g/cm$^3$ for the initial density $(\sigma_\rho/\overline{\rho})_0 \sim 0.87$. A single lower density variance case is run with $(\sigma_\rho/\overline{\rho})_0 \sim 0.3$. One 3D case of higher density variance was run; this case had $R = 120$ (microns) and bead radius $R = 14$ (microns), with 10 higher density (again 4x) beads for each void, yielding $(\sigma_\rho/\overline{\rho})_0 \sim 1.03$.

\section{Shock drive} \label{sec:B}

Here we give more details on the manner in which the shock is launched into the ``working" material where the present study of turbulence generation is focused. As for the construction of the pre-shock density outlined in Appendix \ref{sec:A}, the intent here is not to represent any existing or proposed experiment in detail, but rather to allow for varying conditions to support the present study of turbulence generation. In brief, the five different shock-drive conditions (parameterized here simply by the average shock Mach number) are generated by simulating the ablation of a stacked outer ablator (plastic) and inner pusher (a dense material) by a halfraum radiation drive. The different shock-drive conditions are then created by either scaling the radiation drive temperatures of the halfraum drive (which consists of radiation temperature versus time), or by altering the thicknesses of the ablator and pusher layers.

Thus, while we run without radiation physics when studying the interaction of the shock with the non-uniform density, we run with (60 group) radiation diffusion in the beginning (on order of the first 10 ns) of our simulations, until the shock profile in the right-hand-side of the ``working" material is established. At this point, the radiation physics is turned off (which greatly speeds up the simulations). At all times we run with thermal conduction in the ablator and pusher layers, calculated using the theory of \citet{lee1984}. The heavy inner pusher, which abuts the right hand side of the ``working" material at the start of the simulation (see Fig. \ref{fig:sketch}), serves to enhance the separation of the shock propagation from the motion of the ablator.

Our base halfraum drive follows one inferred from experiments\citep{olson2006} with a peak radiation temperature of $\sim 163$ eV after about 2 ns, extended with a gradually falling tail (see also \citenum{nagel2017}). The ablator is simulated (using tabular opacity, and a quotidian equation of state\citep{more1988}) as a mixture of carbon ($\sim 41.3$ \%), hydrogen ($\sim 57.2$ \%), oxygen ($\sim 0.5$ \%), and germanium ($\sim 1$ \%). The ablator has a density of $\rho \approx 1.13$ g/cm$^3$ while the pusher is taken as density $\rho = 16.7$ g/cm$^3$. At present for the sake of opacity calculations we treat the pusher as copper. This density and opacity combination then does not correspond to a real material but similar shock drives are achievable with self-consistent combinations.

The halfraum drive is applied at the RHS of the simulation domain, which is separated from the ablator by 2 mm of helium gas at a density of $5\times10^{-3}$ g/cm$^3$ (the helium region is truncated in Fig. \ref{fig:sketch} to improve visibility of the rest of the domain). The RHS boundary is free (Lagrangian), allowing for continuing ejection of ablator. No radiation is permitted to leave the domain from the boundaries parallel to the $x$ (shock-propagation) direction, which yields a planar shock.

The five drives we use pair a scaling of the halfraum drive in radiation temperature with a thickness of the ablator and pusher layers. In the format [shock Mach number, peak radiation temperature (eV), pusher thickness (microns), ablator thickness (microns)], these pairings are [12, 195, 15, 200], [15, 228, 20, 250], [18, 228, 15, 100], [20, 228, 10, 200] and [37, 260, 5, 100]. The inclusion of the drive side in the simulation raises the total resolution above the values given in Sec. \ref{sec:approach}, increasing the number of cells in the $x$ (first) direction by approximately 200 -- 400 depending on the case (e.g. the full $R=90$ 3D simulations are $\sim$900x400x400).


\begin{thebibliography}{46}%
\makeatletter
\providecommand \@ifxundefined [1]{%
 \@ifx{#1\undefined}
}%
\providecommand \@ifnum [1]{%
 \ifnum #1\expandafter \@firstoftwo
 \else \expandafter \@secondoftwo
 \fi
}%
\providecommand \@ifx [1]{%
 \ifx #1\expandafter \@firstoftwo
 \else \expandafter \@secondoftwo
 \fi
}%
\providecommand \natexlab [1]{#1}%
\providecommand \enquote  [1]{``#1''}%
\providecommand \bibnamefont  [1]{#1}%
\providecommand \bibfnamefont [1]{#1}%
\providecommand \citenamefont [1]{#1}%
\providecommand \href@noop [0]{\@secondoftwo}%
\providecommand \href [0]{\begingroup \@sanitize@url \@href}%
\providecommand \@href[1]{\@@startlink{#1}\@@href}%
\providecommand \@@href[1]{\endgroup#1\@@endlink}%
\providecommand \@sanitize@url [0]{\catcode `\\12\catcode `\$12\catcode
  `\&12\catcode `\#12\catcode `\^12\catcode `\_12\catcode `\%12\relax}%
\providecommand \@@startlink[1]{}%
\providecommand \@@endlink[0]{}%
\providecommand \url  [0]{\begingroup\@sanitize@url \@url }%
\providecommand \@url [1]{\endgroup\@href {#1}{\urlprefix }}%
\providecommand \urlprefix  [0]{URL }%
\providecommand \Eprint [0]{\href }%
\providecommand \doibase [0]{https://doi.org/}%
\providecommand \selectlanguage [0]{\@gobble}%
\providecommand \bibinfo  [0]{\@secondoftwo}%
\providecommand \bibfield  [0]{\@secondoftwo}%
\providecommand \translation [1]{[#1]}%
\providecommand \BibitemOpen [0]{}%
\providecommand \bibitemStop [0]{}%
\providecommand \bibitemNoStop [0]{.\EOS\space}%
\providecommand \EOS [0]{\spacefactor3000\relax}%
\providecommand \BibitemShut  [1]{\csname bibitem#1\endcsname}%
\let\auto@bib@innerbib\@empty
\bibitem [{\citenamefont {Moore}(1954)}]{moore1954}%
  \BibitemOpen
  \bibfield  {author} {\bibinfo {author} {\bibfnamefont {F.~K.}\ \bibnamefont
  {Moore}},\ }\href@noop {} {\emph {\bibinfo {title} {Unsteady oblique
  interaction of a shock wave with a plane disturbance}}},\ \bibinfo {type}
  {Tech. Rep.}\ (\bibinfo  {institution} {NATIONAL AERONAUTICS AND SPACE
  ADMINISTRATION CLEVELAND OH LEWIS RESEARCH CENTER},\ \bibinfo {year}
  {1954})\BibitemShut {NoStop}%
\bibitem [{\citenamefont {Hesselink}\ and\ \citenamefont
  {Sturtevant}(1988)}]{hesselink1988}%
  \BibitemOpen
  \bibfield  {author} {\bibinfo {author} {\bibfnamefont {L.}~\bibnamefont
  {Hesselink}}\ and\ \bibinfo {author} {\bibfnamefont {B.}~\bibnamefont
  {Sturtevant}},\ }\bibfield  {title} {\bibinfo {title} {Propagation of weak
  shocks through a random medium},\ }\href
  {https://doi.org/10.1017/S0022112088002800} {\bibfield  {journal} {\bibinfo
  {journal} {Journal of Fluid Mechanics}\ }\textbf {\bibinfo {volume} {196}},\
  \bibinfo {pages} {513–553} (\bibinfo {year} {1988})}\BibitemShut {NoStop}%
\bibitem [{\citenamefont {Giacalone}\ and\ \citenamefont
  {Jokipii}(2007)}]{giacalone2007}%
  \BibitemOpen
  \bibfield  {author} {\bibinfo {author} {\bibfnamefont {J.}~\bibnamefont
  {Giacalone}}\ and\ \bibinfo {author} {\bibfnamefont {J.~R.}\ \bibnamefont
  {Jokipii}},\ }\bibfield  {title} {\bibinfo {title} {Magnetic field
  amplification by shocks in turbulent fluids},\ }\href
  {https://doi.org/10.1086/519994} {\bibfield  {journal} {\bibinfo  {journal}
  {The Astrophysical Journal}\ }\textbf {\bibinfo {volume} {663}},\ \bibinfo
  {pages} {L41} (\bibinfo {year} {2007})}\BibitemShut {NoStop}%
\bibitem [{\citenamefont {Inoue}\ \emph {et~al.}(2013)\citenamefont {Inoue},
  \citenamefont {Shimoda}, \citenamefont {Ohira},\ and\ \citenamefont
  {Yamazaki}}]{inoue2013}%
  \BibitemOpen
  \bibfield  {author} {\bibinfo {author} {\bibfnamefont {T.}~\bibnamefont
  {Inoue}}, \bibinfo {author} {\bibfnamefont {J.}~\bibnamefont {Shimoda}},
  \bibinfo {author} {\bibfnamefont {Y.}~\bibnamefont {Ohira}},\ and\ \bibinfo
  {author} {\bibfnamefont {R.}~\bibnamefont {Yamazaki}},\ }\bibfield  {title}
  {\bibinfo {title} {{THE} {ORIGIN} {OF} {RADIALLY} {ALIGNED} {MAGNETIC}
  {FIELDS} {IN} {YOUNG} {SUPERNOVA} {REMNANTS}},\ }\href
  {https://doi.org/10.1088/2041-8205/772/2/l20} {\bibfield  {journal} {\bibinfo
   {journal} {The Astrophysical Journal}\ }\textbf {\bibinfo {volume} {772}},\
  \bibinfo {pages} {L20} (\bibinfo {year} {2013})}\BibitemShut {NoStop}%
\bibitem [{\citenamefont {Banda-Barragán}\ \emph {et~al.}(2021)\citenamefont
  {Banda-Barragán}, \citenamefont {Brüggen}, \citenamefont {Heesen},
  \citenamefont {Scannapieco}, \citenamefont {Cottle}, \citenamefont
  {Federrath},\ and\ \citenamefont {Wagner}}]{banda-barragan2021}%
  \BibitemOpen
  \bibfield  {author} {\bibinfo {author} {\bibfnamefont {W.~E.}\ \bibnamefont
  {Banda-Barragán}}, \bibinfo {author} {\bibfnamefont {M.}~\bibnamefont
  {Brüggen}}, \bibinfo {author} {\bibfnamefont {V.}~\bibnamefont {Heesen}},
  \bibinfo {author} {\bibfnamefont {E.}~\bibnamefont {Scannapieco}}, \bibinfo
  {author} {\bibfnamefont {J.}~\bibnamefont {Cottle}}, \bibinfo {author}
  {\bibfnamefont {C.}~\bibnamefont {Federrath}},\ and\ \bibinfo {author}
  {\bibfnamefont {A.~Y.}\ \bibnamefont {Wagner}},\ }\bibfield  {title}
  {\bibinfo {title} {{Shock–multicloud interactions in galactic outflows –
  II. Radiative fractal clouds and cold gas thermodynamics}},\ }\href
  {https://doi.org/10.1093/mnras/stab1884} {\bibfield  {journal} {\bibinfo
  {journal} {Monthly Notices of the Royal Astronomical Society}\ }\textbf
  {\bibinfo {volume} {506}},\ \bibinfo {pages} {5658} (\bibinfo {year}
  {2021})},\ \Eprint
  {https://arxiv.org/abs/https://academic.oup.com/mnras/article-pdf/506/4/5658/39685437/stab1884.pdf}
  {https://academic.oup.com/mnras/article-pdf/506/4/5658/39685437/stab1884.pdf}
  \BibitemShut {NoStop}%
\bibitem [{\citenamefont {Velikovich}\ \emph {et~al.}(2007)\citenamefont
  {Velikovich}, \citenamefont {Wouchuk}, \citenamefont {Huete Ruiz~de Lira},
  \citenamefont {Metzler}, \citenamefont {Zalesak},\ and\ \citenamefont
  {Schmitt}}]{velikovich2007}%
  \BibitemOpen
  \bibfield  {author} {\bibinfo {author} {\bibfnamefont {A.~L.}\ \bibnamefont
  {Velikovich}}, \bibinfo {author} {\bibfnamefont {J.~G.}\ \bibnamefont
  {Wouchuk}}, \bibinfo {author} {\bibfnamefont {C.}~\bibnamefont {Huete Ruiz~de
  Lira}}, \bibinfo {author} {\bibfnamefont {N.}~\bibnamefont {Metzler}},
  \bibinfo {author} {\bibfnamefont {S.}~\bibnamefont {Zalesak}},\ and\ \bibinfo
  {author} {\bibfnamefont {A.~J.}\ \bibnamefont {Schmitt}},\ }\bibfield
  {title} {\bibinfo {title} {Shock front distortion and richtmyer-meshkov-type
  growth caused by a small preshock nonuniformity},\ }\href
  {https://doi.org/10.1063/1.2745809} {\bibfield  {journal} {\bibinfo
  {journal} {Physics of Plasmas}\ }\textbf {\bibinfo {volume} {14}},\ \bibinfo
  {pages} {072706} (\bibinfo {year} {2007})},\ \Eprint
  {https://arxiv.org/abs/https://doi.org/10.1063/1.2745809}
  {https://doi.org/10.1063/1.2745809} \BibitemShut {NoStop}%
\bibitem [{\citenamefont {Velikovich}\ \emph {et~al.}(2012)\citenamefont
  {Velikovich}, \citenamefont {Huete},\ and\ \citenamefont
  {Wouchuk}}]{velikovich2012}%
  \BibitemOpen
  \bibfield  {author} {\bibinfo {author} {\bibfnamefont {A.~L.}\ \bibnamefont
  {Velikovich}}, \bibinfo {author} {\bibfnamefont {C.}~\bibnamefont {Huete}},\
  and\ \bibinfo {author} {\bibfnamefont {J.~G.}\ \bibnamefont {Wouchuk}},\
  }\bibfield  {title} {\bibinfo {title} {Effect of shock-generated turbulence
  on the hugoniot jump conditions},\ }\href
  {https://doi.org/10.1103/PhysRevE.85.016301} {\bibfield  {journal} {\bibinfo
  {journal} {Phys. Rev. E}\ }\textbf {\bibinfo {volume} {85}},\ \bibinfo
  {pages} {016301} (\bibinfo {year} {2012})}\BibitemShut {NoStop}%
\bibitem [{\citenamefont {Ali}\ \emph {et~al.}(2018)\citenamefont {Ali},
  \citenamefont {Celliers}, \citenamefont {Haan}, \citenamefont {Boehly},
  \citenamefont {Whiting}, \citenamefont {Baxamusa}, \citenamefont {Reynolds},
  \citenamefont {Johnson}, \citenamefont {Hughes}, \citenamefont {Watson},
  \citenamefont {Huang}, \citenamefont {Biener}, \citenamefont {Engelhorn},
  \citenamefont {Smalyuk},\ and\ \citenamefont {Landen}}]{ali2018}%
  \BibitemOpen
  \bibfield  {author} {\bibinfo {author} {\bibfnamefont {S.~J.}\ \bibnamefont
  {Ali}}, \bibinfo {author} {\bibfnamefont {P.~M.}\ \bibnamefont {Celliers}},
  \bibinfo {author} {\bibfnamefont {S.}~\bibnamefont {Haan}}, \bibinfo {author}
  {\bibfnamefont {T.~R.}\ \bibnamefont {Boehly}}, \bibinfo {author}
  {\bibfnamefont {N.}~\bibnamefont {Whiting}}, \bibinfo {author} {\bibfnamefont
  {S.~H.}\ \bibnamefont {Baxamusa}}, \bibinfo {author} {\bibfnamefont
  {H.}~\bibnamefont {Reynolds}}, \bibinfo {author} {\bibfnamefont {M.~A.}\
  \bibnamefont {Johnson}}, \bibinfo {author} {\bibfnamefont {J.~D.}\
  \bibnamefont {Hughes}}, \bibinfo {author} {\bibfnamefont {B.}~\bibnamefont
  {Watson}}, \bibinfo {author} {\bibfnamefont {H.}~\bibnamefont {Huang}},
  \bibinfo {author} {\bibfnamefont {J.}~\bibnamefont {Biener}}, \bibinfo
  {author} {\bibfnamefont {K.}~\bibnamefont {Engelhorn}}, \bibinfo {author}
  {\bibfnamefont {V.~A.}\ \bibnamefont {Smalyuk}},\ and\ \bibinfo {author}
  {\bibfnamefont {O.~L.}\ \bibnamefont {Landen}},\ }\bibfield  {title}
  {\bibinfo {title} {Probing the seeding of hydrodynamic instabilities from
  nonuniformities in ablator materials using 2d velocimetry},\ }\href
  {https://doi.org/10.1063/1.5047943} {\bibfield  {journal} {\bibinfo
  {journal} {Physics of Plasmas}\ }\textbf {\bibinfo {volume} {25}},\ \bibinfo
  {pages} {092708} (\bibinfo {year} {2018})},\ \Eprint
  {https://arxiv.org/abs/https://doi.org/10.1063/1.5047943}
  {https://doi.org/10.1063/1.5047943} \BibitemShut {NoStop}%
\bibitem [{\citenamefont {Remington}\ \emph {et~al.}(2006)\citenamefont
  {Remington}, \citenamefont {Drake},\ and\ \citenamefont
  {Ryutov}}]{remington2006}%
  \BibitemOpen
  \bibfield  {author} {\bibinfo {author} {\bibfnamefont {B.~A.}\ \bibnamefont
  {Remington}}, \bibinfo {author} {\bibfnamefont {R.~P.}\ \bibnamefont
  {Drake}},\ and\ \bibinfo {author} {\bibfnamefont {D.~D.}\ \bibnamefont
  {Ryutov}},\ }\bibfield  {title} {\bibinfo {title} {Experimental astrophysics
  with high power lasers and $z$ pinches},\ }\href
  {https://doi.org/10.1103/RevModPhys.78.755} {\bibfield  {journal} {\bibinfo
  {journal} {Rev. Mod. Phys.}\ }\textbf {\bibinfo {volume} {78}},\ \bibinfo
  {pages} {755} (\bibinfo {year} {2006})}\BibitemShut {NoStop}%
\bibitem [{\citenamefont {Federrath}(2018)}]{federrath2018}%
  \BibitemOpen
  \bibfield  {author} {\bibinfo {author} {\bibfnamefont {C.}~\bibnamefont
  {Federrath}},\ }\bibfield  {title} {\bibinfo {title} {The turbulent formation
  of stars},\ }\href {https://doi.org/10.1063/PT.3.3947} {\bibfield  {journal}
  {\bibinfo  {journal} {Physics Today}\ }\textbf {\bibinfo {volume} {71}},\
  \bibinfo {pages} {38} (\bibinfo {year} {2018})},\ \Eprint
  {https://arxiv.org/abs/https://doi.org/10.1063/PT.3.3947}
  {https://doi.org/10.1063/PT.3.3947} \BibitemShut {NoStop}%
\bibitem [{\citenamefont {Meinecke}\ \emph {et~al.}(2015)\citenamefont
  {Meinecke}, \citenamefont {Tzeferacos}, \citenamefont {Bell}, \citenamefont
  {Bingham}, \citenamefont {Clarke}, \citenamefont {Churazov}, \citenamefont
  {Crowston}, \citenamefont {Doyle}, \citenamefont {Drake}, \citenamefont
  {Heathcote} \emph {et~al.}}]{meinecke2015}%
  \BibitemOpen
  \bibfield  {author} {\bibinfo {author} {\bibfnamefont {J.}~\bibnamefont
  {Meinecke}}, \bibinfo {author} {\bibfnamefont {P.}~\bibnamefont
  {Tzeferacos}}, \bibinfo {author} {\bibfnamefont {A.}~\bibnamefont {Bell}},
  \bibinfo {author} {\bibfnamefont {R.}~\bibnamefont {Bingham}}, \bibinfo
  {author} {\bibfnamefont {R.}~\bibnamefont {Clarke}}, \bibinfo {author}
  {\bibfnamefont {E.}~\bibnamefont {Churazov}}, \bibinfo {author}
  {\bibfnamefont {R.}~\bibnamefont {Crowston}}, \bibinfo {author}
  {\bibfnamefont {H.}~\bibnamefont {Doyle}}, \bibinfo {author} {\bibfnamefont
  {R.~P.}\ \bibnamefont {Drake}}, \bibinfo {author} {\bibfnamefont
  {R.}~\bibnamefont {Heathcote}}, \emph {et~al.},\ }\bibfield  {title}
  {\bibinfo {title} {Developed turbulence and nonlinear amplification of
  magnetic fields in laboratory and astrophysical plasmas},\ }\href
  {https://doi.org/10.1073/pnas.1502079112} {\bibfield  {journal} {\bibinfo
  {journal} {Proceedings of the National Academy of Sciences}\ }\textbf
  {\bibinfo {volume} {112}},\ \bibinfo {pages} {8211} (\bibinfo {year}
  {2015})}\BibitemShut {NoStop}%
\bibitem [{\citenamefont {Tzeferacos}\ \emph {et~al.}(2018)\citenamefont
  {Tzeferacos}, \citenamefont {Rigby}, \citenamefont {Bott}, \citenamefont
  {Bell}, \citenamefont {Bingham}, \citenamefont {Casner}, \citenamefont
  {Cattaneo}, \citenamefont {Churazov}, \citenamefont {Emig}, \citenamefont
  {Fiuza} \emph {et~al.}}]{tzeferacos2018}%
  \BibitemOpen
  \bibfield  {author} {\bibinfo {author} {\bibfnamefont {P.}~\bibnamefont
  {Tzeferacos}}, \bibinfo {author} {\bibfnamefont {A.}~\bibnamefont {Rigby}},
  \bibinfo {author} {\bibfnamefont {A.}~\bibnamefont {Bott}}, \bibinfo {author}
  {\bibfnamefont {A.}~\bibnamefont {Bell}}, \bibinfo {author} {\bibfnamefont
  {R.}~\bibnamefont {Bingham}}, \bibinfo {author} {\bibfnamefont
  {A.}~\bibnamefont {Casner}}, \bibinfo {author} {\bibfnamefont
  {F.}~\bibnamefont {Cattaneo}}, \bibinfo {author} {\bibfnamefont
  {E.}~\bibnamefont {Churazov}}, \bibinfo {author} {\bibfnamefont
  {J.}~\bibnamefont {Emig}}, \bibinfo {author} {\bibfnamefont {F.}~\bibnamefont
  {Fiuza}}, \emph {et~al.},\ }\bibfield  {title} {\bibinfo {title} {Laboratory
  evidence of dynamo amplification of magnetic fields in a turbulent plasma},\
  }\href {https://doi.org/10.1038/s41467-018-02953-2} {\bibfield  {journal}
  {\bibinfo  {journal} {Nature Communications}\ }\textbf {\bibinfo {volume}
  {9}},\ \bibinfo {pages} {1} (\bibinfo {year} {2018})}\BibitemShut {NoStop}%
\bibitem [{\citenamefont {White}\ \emph {et~al.}(2019)\citenamefont {White},
  \citenamefont {Oliver}, \citenamefont {Mabey}, \citenamefont
  {K{\"u}hn-Kauffeldt}, \citenamefont {Bott}, \citenamefont {D{\"o}hl},
  \citenamefont {Bell}, \citenamefont {Bingham}, \citenamefont {Clarke},
  \citenamefont {Foster} \emph {et~al.}}]{white2019}%
  \BibitemOpen
  \bibfield  {author} {\bibinfo {author} {\bibfnamefont {T.}~\bibnamefont
  {White}}, \bibinfo {author} {\bibfnamefont {M.}~\bibnamefont {Oliver}},
  \bibinfo {author} {\bibfnamefont {P.}~\bibnamefont {Mabey}}, \bibinfo
  {author} {\bibfnamefont {M.}~\bibnamefont {K{\"u}hn-Kauffeldt}}, \bibinfo
  {author} {\bibfnamefont {A.}~\bibnamefont {Bott}}, \bibinfo {author}
  {\bibfnamefont {L.}~\bibnamefont {D{\"o}hl}}, \bibinfo {author}
  {\bibfnamefont {A.}~\bibnamefont {Bell}}, \bibinfo {author} {\bibfnamefont
  {R.}~\bibnamefont {Bingham}}, \bibinfo {author} {\bibfnamefont
  {R.}~\bibnamefont {Clarke}}, \bibinfo {author} {\bibfnamefont
  {J.}~\bibnamefont {Foster}}, \emph {et~al.},\ }\bibfield  {title} {\bibinfo
  {title} {Supersonic plasma turbulence in the laboratory},\ }\href
  {https://doi.org/10.1038/s41467-019-09498-y} {\bibfield  {journal} {\bibinfo
  {journal} {Nature communications}\ }\textbf {\bibinfo {volume} {10}},\
  \bibinfo {pages} {1} (\bibinfo {year} {2019})}\BibitemShut {NoStop}%
\bibitem [{\citenamefont {Nagel}\ \emph {et~al.}(2017)\citenamefont {Nagel},
  \citenamefont {Raman}, \citenamefont {Huntington}, \citenamefont {MacLaren},
  \citenamefont {Wang}, \citenamefont {Barrios}, \citenamefont {Baumann},
  \citenamefont {Bender}, \citenamefont {Benedetti}, \citenamefont {Doane},
  \citenamefont {Felker}, \citenamefont {Fitzsimmons}, \citenamefont {Flippo},
  \citenamefont {Holder}, \citenamefont {Kaczala}, \citenamefont {Perry},
  \citenamefont {Seugling}, \citenamefont {Savage},\ and\ \citenamefont
  {Zhou}}]{nagel2017}%
  \BibitemOpen
  \bibfield  {author} {\bibinfo {author} {\bibfnamefont {S.~R.}\ \bibnamefont
  {Nagel}}, \bibinfo {author} {\bibfnamefont {K.~S.}\ \bibnamefont {Raman}},
  \bibinfo {author} {\bibfnamefont {C.~M.}\ \bibnamefont {Huntington}},
  \bibinfo {author} {\bibfnamefont {S.~A.}\ \bibnamefont {MacLaren}}, \bibinfo
  {author} {\bibfnamefont {P.}~\bibnamefont {Wang}}, \bibinfo {author}
  {\bibfnamefont {M.~A.}\ \bibnamefont {Barrios}}, \bibinfo {author}
  {\bibfnamefont {T.}~\bibnamefont {Baumann}}, \bibinfo {author} {\bibfnamefont
  {J.~D.}\ \bibnamefont {Bender}}, \bibinfo {author} {\bibfnamefont {L.~R.}\
  \bibnamefont {Benedetti}}, \bibinfo {author} {\bibfnamefont {D.~M.}\
  \bibnamefont {Doane}}, \bibinfo {author} {\bibfnamefont {S.}~\bibnamefont
  {Felker}}, \bibinfo {author} {\bibfnamefont {P.}~\bibnamefont {Fitzsimmons}},
  \bibinfo {author} {\bibfnamefont {K.~A.}\ \bibnamefont {Flippo}}, \bibinfo
  {author} {\bibfnamefont {J.~P.}\ \bibnamefont {Holder}}, \bibinfo {author}
  {\bibfnamefont {D.~N.}\ \bibnamefont {Kaczala}}, \bibinfo {author}
  {\bibfnamefont {T.~S.}\ \bibnamefont {Perry}}, \bibinfo {author}
  {\bibfnamefont {R.~M.}\ \bibnamefont {Seugling}}, \bibinfo {author}
  {\bibfnamefont {L.}~\bibnamefont {Savage}},\ and\ \bibinfo {author}
  {\bibfnamefont {Y.}~\bibnamefont {Zhou}},\ }\bibfield  {title} {\bibinfo
  {title} {A platform for studying the rayleigh–taylor and
  richtmyer–meshkov instabilities in a planar geometry at high energy density
  at the national ignition facility},\ }\href
  {https://doi.org/10.1063/1.4985312} {\bibfield  {journal} {\bibinfo
  {journal} {Physics of Plasmas}\ }\textbf {\bibinfo {volume} {24}},\ \bibinfo
  {pages} {072704} (\bibinfo {year} {2017})},\ \Eprint
  {https://arxiv.org/abs/https://doi.org/10.1063/1.4985312}
  {https://doi.org/10.1063/1.4985312} \BibitemShut {NoStop}%
\bibitem [{\citenamefont {Hamilton}\ \emph {et~al.}(2016)\citenamefont
  {Hamilton}, \citenamefont {Lee},\ and\ \citenamefont
  {Parra-Vasquez}}]{hamilton2016}%
  \BibitemOpen
  \bibfield  {author} {\bibinfo {author} {\bibfnamefont {C.~E.}\ \bibnamefont
  {Hamilton}}, \bibinfo {author} {\bibfnamefont {M.~N.}\ \bibnamefont {Lee}},\
  and\ \bibinfo {author} {\bibfnamefont {A.~N.~G.}\ \bibnamefont
  {Parra-Vasquez}},\ }\bibfield  {title} {\bibinfo {title} {Development of
  hierarchical, tunable pore size polymer foams for icf targets},\ }\href
  {https://doi.org/10.13182/FST15-227} {\bibfield  {journal} {\bibinfo
  {journal} {Fusion Science and Technology}\ }\textbf {\bibinfo {volume}
  {70}},\ \bibinfo {pages} {226} (\bibinfo {year} {2016})},\ \Eprint
  {https://arxiv.org/abs/https://doi.org/10.13182/FST15-227}
  {https://doi.org/10.13182/FST15-227} \BibitemShut {NoStop}%
\bibitem [{\citenamefont {Saha}\ \emph {et~al.}(2018)\citenamefont {Saha},
  \citenamefont {Oakdale}, \citenamefont {Cuadra}, \citenamefont {Divin},
  \citenamefont {Ye}, \citenamefont {Forien}, \citenamefont {Bayu~Aji},
  \citenamefont {Biener},\ and\ \citenamefont {Smith}}]{saha2018}%
  \BibitemOpen
  \bibfield  {author} {\bibinfo {author} {\bibfnamefont {S.~K.}\ \bibnamefont
  {Saha}}, \bibinfo {author} {\bibfnamefont {J.~S.}\ \bibnamefont {Oakdale}},
  \bibinfo {author} {\bibfnamefont {J.~A.}\ \bibnamefont {Cuadra}}, \bibinfo
  {author} {\bibfnamefont {C.}~\bibnamefont {Divin}}, \bibinfo {author}
  {\bibfnamefont {J.}~\bibnamefont {Ye}}, \bibinfo {author} {\bibfnamefont
  {J.-B.}\ \bibnamefont {Forien}}, \bibinfo {author} {\bibfnamefont {L.~B.}\
  \bibnamefont {Bayu~Aji}}, \bibinfo {author} {\bibfnamefont {J.}~\bibnamefont
  {Biener}},\ and\ \bibinfo {author} {\bibfnamefont {W.~L.}\ \bibnamefont
  {Smith}},\ }\bibfield  {title} {\bibinfo {title} {Radiopaque resists for
  two-photon lithography to enable submicron 3d imaging of polymer parts via
  x-ray computed tomography},\ }\href {https://doi.org/10.1021/acsami.7b12654}
  {\bibfield  {journal} {\bibinfo  {journal} {ACS Applied Materials \&
  Interfaces}\ }\textbf {\bibinfo {volume} {10}},\ \bibinfo {pages} {1164}
  (\bibinfo {year} {2018})},\ \bibinfo {note} {pMID: 29171264},\ \Eprint
  {https://arxiv.org/abs/https://doi.org/10.1021/acsami.7b12654}
  {https://doi.org/10.1021/acsami.7b12654} \BibitemShut {NoStop}%
\bibitem [{\citenamefont {Murphy}\ \emph {et~al.}(2016)\citenamefont {Murphy},
  \citenamefont {Douglas}, \citenamefont {Fincke}, \citenamefont {Olson},
  \citenamefont {Cobble}, \citenamefont {Haines}, \citenamefont {Hamilton},
  \citenamefont {Lee}, \citenamefont {Oertel}, \citenamefont {Parra-Vasquez}
  \emph {et~al.}}]{murphy2016}%
  \BibitemOpen
  \bibfield  {author} {\bibinfo {author} {\bibfnamefont {T.~J.}\ \bibnamefont
  {Murphy}}, \bibinfo {author} {\bibfnamefont {M.~R.}\ \bibnamefont {Douglas}},
  \bibinfo {author} {\bibfnamefont {J.~R.}\ \bibnamefont {Fincke}}, \bibinfo
  {author} {\bibfnamefont {R.}~\bibnamefont {Olson}}, \bibinfo {author}
  {\bibfnamefont {J.~A.}\ \bibnamefont {Cobble}}, \bibinfo {author}
  {\bibfnamefont {B.~M.}\ \bibnamefont {Haines}}, \bibinfo {author}
  {\bibfnamefont {C.}~\bibnamefont {Hamilton}}, \bibinfo {author}
  {\bibfnamefont {M.~N.}\ \bibnamefont {Lee}}, \bibinfo {author} {\bibfnamefont
  {J.~A.}\ \bibnamefont {Oertel}}, \bibinfo {author} {\bibfnamefont
  {N.}~\bibnamefont {Parra-Vasquez}}, \emph {et~al.},\ }\bibfield  {title}
  {\bibinfo {title} {Progress in the development of the marble platform for
  studying thermonuclear burn in the presence of heterogeneous mix on omega and
  the national ignition facility},\ }in\ \href@noop {} {\emph {\bibinfo
  {booktitle} {Journal of Physics: Conference Series}}},\ Vol.\ \bibinfo
  {volume} {717}\ (\bibinfo {organization} {IOP Publishing},\ \bibinfo {year}
  {2016})\ p.\ \bibinfo {pages} {012072}\BibitemShut {NoStop}%
\bibitem [{\citenamefont {Haines}\ \emph {et~al.}(2020)\citenamefont {Haines},
  \citenamefont {Shah}, \citenamefont {Smidt}, \citenamefont {Albright},
  \citenamefont {Cardenas}, \citenamefont {Douglas}, \citenamefont {Forrest},
  \citenamefont {Glebov}, \citenamefont {Gunderson}, \citenamefont {Hamilton}
  \emph {et~al.}}]{haines2020}%
  \BibitemOpen
  \bibfield  {author} {\bibinfo {author} {\bibfnamefont {B.~M.}\ \bibnamefont
  {Haines}}, \bibinfo {author} {\bibfnamefont {R.~C.}\ \bibnamefont {Shah}},
  \bibinfo {author} {\bibfnamefont {J.~M.}\ \bibnamefont {Smidt}}, \bibinfo
  {author} {\bibfnamefont {B.~J.}\ \bibnamefont {Albright}}, \bibinfo {author}
  {\bibfnamefont {T.}~\bibnamefont {Cardenas}}, \bibinfo {author}
  {\bibfnamefont {M.~R.}\ \bibnamefont {Douglas}}, \bibinfo {author}
  {\bibfnamefont {C.}~\bibnamefont {Forrest}}, \bibinfo {author} {\bibfnamefont
  {V.~Y.}\ \bibnamefont {Glebov}}, \bibinfo {author} {\bibfnamefont {M.~A.}\
  \bibnamefont {Gunderson}}, \bibinfo {author} {\bibfnamefont {C.~E.}\
  \bibnamefont {Hamilton}}, \emph {et~al.},\ }\bibfield  {title} {\bibinfo
  {title} {Observation of persistent species temperature separation in inertial
  confinement fusion mixtures},\ }\href@noop {} {\bibfield  {journal} {\bibinfo
   {journal} {Nature communications}\ }\textbf {\bibinfo {volume} {11}},\
  \bibinfo {pages} {1} (\bibinfo {year} {2020})}\BibitemShut {NoStop}%
\bibitem [{\citenamefont {Ribner}(1955)}]{ribner1955}%
  \BibitemOpen
  \bibfield  {author} {\bibinfo {author} {\bibfnamefont {H.~S.}\ \bibnamefont
  {Ribner}},\ }\href@noop {} {\emph {\bibinfo {title} {Shock-Turbulence
  Interaction and the Generation of Noise}}},\ \bibinfo {type} {Tech. Rep.}\
  (\bibinfo  {institution} {NATIONAL AERONAUTICS AND SPACE ADMINISTRATION
  CLEVELAND OH LEWIS RESEARCH CENTER},\ \bibinfo {year} {1955})\BibitemShut
  {NoStop}%
\bibitem [{\citenamefont {Ribner}(1987)}]{ribner1987}%
  \BibitemOpen
  \bibfield  {author} {\bibinfo {author} {\bibfnamefont {H.~S.}\ \bibnamefont
  {Ribner}},\ }\bibfield  {title} {\bibinfo {title} {Spectra of noise and
  amplified turbulence emanating from shock-turbulence interaction},\ }\href
  {https://doi.org/10.2514/3.9642} {\bibfield  {journal} {\bibinfo  {journal}
  {AIAA Journal}\ }\textbf {\bibinfo {volume} {25}},\ \bibinfo {pages} {436}
  (\bibinfo {year} {1987})},\ \Eprint
  {https://arxiv.org/abs/https://doi.org/10.2514/3.9642}
  {https://doi.org/10.2514/3.9642} \BibitemShut {NoStop}%
\bibitem [{\citenamefont {Durbin}\ and\ \citenamefont
  {Zeman}(1992)}]{durbin1992}%
  \BibitemOpen
  \bibfield  {author} {\bibinfo {author} {\bibfnamefont {P.~A.}\ \bibnamefont
  {Durbin}}\ and\ \bibinfo {author} {\bibfnamefont {O.}~\bibnamefont {Zeman}},\
  }\bibfield  {title} {\bibinfo {title} {Rapid distortion theory for
  homogeneous compressed turbulence with application to modelling},\ }\href
  {https://doi.org/10.1017/S0022112092002404} {\bibfield  {journal} {\bibinfo
  {journal} {Journal of Fluid Mechanics}\ }\textbf {\bibinfo {volume} {242}},\
  \bibinfo {pages} {349} (\bibinfo {year} {1992})}\BibitemShut {NoStop}%
\bibitem [{\citenamefont {Mahesh}\ \emph {et~al.}(1995)\citenamefont {Mahesh},
  \citenamefont {Lee}, \citenamefont {Lele},\ and\ \citenamefont
  {Moin}}]{mahesh1995}%
  \BibitemOpen
  \bibfield  {author} {\bibinfo {author} {\bibfnamefont {K.}~\bibnamefont
  {Mahesh}}, \bibinfo {author} {\bibfnamefont {S.}~\bibnamefont {Lee}},
  \bibinfo {author} {\bibfnamefont {S.~K.}\ \bibnamefont {Lele}},\ and\
  \bibinfo {author} {\bibfnamefont {P.}~\bibnamefont {Moin}},\ }\bibfield
  {title} {\bibinfo {title} {The interaction of an isotropic field of acoustic
  waves with a shock wave},\ }\href {https://doi.org/10.1017/S0022112095003739}
  {\bibfield  {journal} {\bibinfo  {journal} {Journal of Fluid Mechanics}\
  }\textbf {\bibinfo {volume} {300}},\ \bibinfo {pages} {383–407} (\bibinfo
  {year} {1995})}\BibitemShut {NoStop}%
\bibitem [{\citenamefont {Lele}(1992)}]{lele1992}%
  \BibitemOpen
  \bibfield  {author} {\bibinfo {author} {\bibfnamefont {S.~K.}\ \bibnamefont
  {Lele}},\ }\bibfield  {title} {\bibinfo {title} {Shock‐jump relations in a
  turbulent flow},\ }\href {https://doi.org/10.1063/1.858343} {\bibfield
  {journal} {\bibinfo  {journal} {Physics of Fluids A: Fluid Dynamics}\
  }\textbf {\bibinfo {volume} {4}},\ \bibinfo {pages} {2900} (\bibinfo {year}
  {1992})},\ \Eprint {https://arxiv.org/abs/https://doi.org/10.1063/1.858343}
  {https://doi.org/10.1063/1.858343} \BibitemShut {NoStop}%
\bibitem [{\citenamefont {Zank}\ \emph {et~al.}(2002)\citenamefont {Zank},
  \citenamefont {Zhou}, \citenamefont {Matthaeus},\ and\ \citenamefont
  {Rice}}]{zank2002}%
  \BibitemOpen
  \bibfield  {author} {\bibinfo {author} {\bibfnamefont {G.~P.}\ \bibnamefont
  {Zank}}, \bibinfo {author} {\bibfnamefont {Y.}~\bibnamefont {Zhou}}, \bibinfo
  {author} {\bibfnamefont {W.~H.}\ \bibnamefont {Matthaeus}},\ and\ \bibinfo
  {author} {\bibfnamefont {W.~K.~M.}\ \bibnamefont {Rice}},\ }\bibfield
  {title} {\bibinfo {title} {The interaction of turbulence with shock waves: A
  basic model},\ }\href {https://doi.org/10.1063/1.1507772} {\bibfield
  {journal} {\bibinfo  {journal} {Physics of Fluids}\ }\textbf {\bibinfo
  {volume} {14}},\ \bibinfo {pages} {3766} (\bibinfo {year} {2002})},\ \Eprint
  {https://arxiv.org/abs/https://doi.org/10.1063/1.1507772}
  {https://doi.org/10.1063/1.1507772} \BibitemShut {NoStop}%
\bibitem [{\citenamefont {Huete Ruiz~de Lira}\ \emph
  {et~al.}(2011)\citenamefont {Huete Ruiz~de Lira}, \citenamefont
  {Velikovich},\ and\ \citenamefont {Wouchuk}}]{huete2011}%
  \BibitemOpen
  \bibfield  {author} {\bibinfo {author} {\bibfnamefont {C.}~\bibnamefont
  {Huete Ruiz~de Lira}}, \bibinfo {author} {\bibfnamefont {A.~L.}\ \bibnamefont
  {Velikovich}},\ and\ \bibinfo {author} {\bibfnamefont {J.~G.}\ \bibnamefont
  {Wouchuk}},\ }\bibfield  {title} {\bibinfo {title} {Analytical linear theory
  for the interaction of a planar shock wave with a two- or three-dimensional
  random isotropic density field},\ }\href
  {https://doi.org/10.1103/PhysRevE.83.056320} {\bibfield  {journal} {\bibinfo
  {journal} {Phys. Rev. E}\ }\textbf {\bibinfo {volume} {83}},\ \bibinfo
  {pages} {056320} (\bibinfo {year} {2011})}\BibitemShut {NoStop}%
\bibitem [{\citenamefont {Kitamura}\ \emph {et~al.}(2016)\citenamefont
  {Kitamura}, \citenamefont {Nagata}, \citenamefont {Sakai}, \citenamefont
  {Sasoh},\ and\ \citenamefont {Ito}}]{kitamura2016}%
  \BibitemOpen
  \bibfield  {author} {\bibinfo {author} {\bibfnamefont {T.}~\bibnamefont
  {Kitamura}}, \bibinfo {author} {\bibfnamefont {K.}~\bibnamefont {Nagata}},
  \bibinfo {author} {\bibfnamefont {Y.}~\bibnamefont {Sakai}}, \bibinfo
  {author} {\bibfnamefont {A.}~\bibnamefont {Sasoh}},\ and\ \bibinfo {author}
  {\bibfnamefont {Y.}~\bibnamefont {Ito}},\ }\bibfield  {title} {\bibinfo
  {title} {Rapid distortion theory analysis on the interaction between
  homogeneous turbulence and a planar shock wave},\ }\href
  {https://doi.org/10.1017/jfm.2016.313} {\bibfield  {journal} {\bibinfo
  {journal} {Journal of Fluid Mechanics}\ }\textbf {\bibinfo {volume} {802}},\
  \bibinfo {pages} {108–146} (\bibinfo {year} {2016})}\BibitemShut {NoStop}%
\bibitem [{\citenamefont {Chen}\ and\ \citenamefont {Donzis}(2019)}]{chen2019}%
  \BibitemOpen
  \bibfield  {author} {\bibinfo {author} {\bibfnamefont {C.~H.}\ \bibnamefont
  {Chen}}\ and\ \bibinfo {author} {\bibfnamefont {D.~A.}\ \bibnamefont
  {Donzis}},\ }\bibfield  {title} {\bibinfo {title} {Shock–turbulence
  interactions at high turbulence intensities},\ }\href
  {https://doi.org/10.1017/jfm.2019.248} {\bibfield  {journal} {\bibinfo
  {journal} {Journal of Fluid Mechanics}\ }\textbf {\bibinfo {volume} {870}},\
  \bibinfo {pages} {813–847} (\bibinfo {year} {2019})}\BibitemShut {NoStop}%
\bibitem [{\citenamefont {Donzis}(2012)}]{donzis2012}%
  \BibitemOpen
  \bibfield  {author} {\bibinfo {author} {\bibfnamefont {D.~A.}\ \bibnamefont
  {Donzis}},\ }\bibfield  {title} {\bibinfo {title} {Shock structure in
  shock-turbulence interactions},\ }\href {https://doi.org/10.1063/1.4772064}
  {\bibfield  {journal} {\bibinfo  {journal} {Physics of Fluids}\ }\textbf
  {\bibinfo {volume} {24}},\ \bibinfo {pages} {126101} (\bibinfo {year}
  {2012})},\ \Eprint {https://arxiv.org/abs/https://doi.org/10.1063/1.4772064}
  {https://doi.org/10.1063/1.4772064} \BibitemShut {NoStop}%
\bibitem [{\citenamefont {Richtmyer}(1960)}]{richtmyer1960}%
  \BibitemOpen
  \bibfield  {author} {\bibinfo {author} {\bibfnamefont {R.~D.}\ \bibnamefont
  {Richtmyer}},\ }\bibfield  {title} {\bibinfo {title} {Taylor instability in
  shock acceleration of compressible fluids},\ }\href
  {https://doi.org/https://doi.org/10.1002/cpa.3160130207} {\bibfield
  {journal} {\bibinfo  {journal} {Communications on Pure and Applied
  Mathematics}\ }\textbf {\bibinfo {volume} {13}},\ \bibinfo {pages} {297}
  (\bibinfo {year} {1960})},\ \Eprint
  {https://arxiv.org/abs/https://onlinelibrary.wiley.com/doi/pdf/10.1002/cpa.3160130207}
  {https://onlinelibrary.wiley.com/doi/pdf/10.1002/cpa.3160130207} \BibitemShut
  {NoStop}%
\bibitem [{\citenamefont {Marinak}\ \emph {et~al.}(2001)\citenamefont
  {Marinak}, \citenamefont {Kerbel}, \citenamefont {Gentile}, \citenamefont
  {Jones}, \citenamefont {Munro}, \citenamefont {Pollaine}, \citenamefont
  {Dittrich},\ and\ \citenamefont {Haan}}]{marinak2001}%
  \BibitemOpen
  \bibfield  {author} {\bibinfo {author} {\bibfnamefont {M.~M.}\ \bibnamefont
  {Marinak}}, \bibinfo {author} {\bibfnamefont {G.~D.}\ \bibnamefont {Kerbel}},
  \bibinfo {author} {\bibfnamefont {N.~A.}\ \bibnamefont {Gentile}}, \bibinfo
  {author} {\bibfnamefont {O.}~\bibnamefont {Jones}}, \bibinfo {author}
  {\bibfnamefont {D.}~\bibnamefont {Munro}}, \bibinfo {author} {\bibfnamefont
  {S.}~\bibnamefont {Pollaine}}, \bibinfo {author} {\bibfnamefont {T.~R.}\
  \bibnamefont {Dittrich}},\ and\ \bibinfo {author} {\bibfnamefont {S.~W.}\
  \bibnamefont {Haan}},\ }\bibfield  {title} {\bibinfo {title}
  {Three-dimensional hydra simulations of national ignition facility targets},\
  }\href {https://doi.org/10.1063/1.1356740} {\bibfield  {journal} {\bibinfo
  {journal} {Physics of Plasmas}\ }\textbf {\bibinfo {volume} {8}},\ \bibinfo
  {pages} {2275} (\bibinfo {year} {2001})},\ \Eprint
  {https://arxiv.org/abs/https://doi.org/10.1063/1.1356740}
  {https://doi.org/10.1063/1.1356740} \BibitemShut {NoStop}%
\bibitem [{\citenamefont {Sagaut}(2006)}]{sagaut2006}%
  \BibitemOpen
  \bibfield  {author} {\bibinfo {author} {\bibfnamefont {P.}~\bibnamefont
  {Sagaut}},\ }\href@noop {} {\emph {\bibinfo {title} {Large eddy simulation
  for incompressible flows: an introduction}}}\ (\bibinfo  {publisher}
  {Springer Science \& Business Media},\ \bibinfo {year} {2006})\BibitemShut
  {NoStop}%
\bibitem [{\citenamefont {{Kitsionas, S.}}\ \emph {et~al.}(2009)\citenamefont
  {{Kitsionas, S.}}, \citenamefont {{Federrath, C.}}, \citenamefont {{Klessen,
  R. S.}}, \citenamefont {{Schmidt, W.}}, \citenamefont {{Price, D. J.}},
  \citenamefont {{Dursi, L. J.}}, \citenamefont {{Gritschneder, M.}},
  \citenamefont {{Walch, S.}}, \citenamefont {{Piontek, R.}}, \citenamefont
  {{Kim, J.}}, \citenamefont {{Jappsen, A.-K.}}, \citenamefont {{Ciecielag,
  P.}},\ and\ \citenamefont {{Mac Low, M.-M.}}}]{kitsionas2009}%
  \BibitemOpen
  \bibfield  {author} {\bibinfo {author} {\bibnamefont {{Kitsionas, S.}}},
  \bibinfo {author} {\bibnamefont {{Federrath, C.}}}, \bibinfo {author}
  {\bibnamefont {{Klessen, R. S.}}}, \bibinfo {author} {\bibnamefont {{Schmidt,
  W.}}}, \bibinfo {author} {\bibnamefont {{Price, D. J.}}}, \bibinfo {author}
  {\bibnamefont {{Dursi, L. J.}}}, \bibinfo {author} {\bibnamefont
  {{Gritschneder, M.}}}, \bibinfo {author} {\bibnamefont {{Walch, S.}}},
  \bibinfo {author} {\bibnamefont {{Piontek, R.}}}, \bibinfo {author}
  {\bibnamefont {{Kim, J.}}}, \bibinfo {author} {\bibnamefont {{Jappsen,
  A.-K.}}}, \bibinfo {author} {\bibnamefont {{Ciecielag, P.}}},\ and\ \bibinfo
  {author} {\bibnamefont {{Mac Low, M.-M.}}},\ }\bibfield  {title} {\bibinfo
  {title} {Algorithmic comparisons of decaying, isothermal, supersonic
  turbulence*},\ }\href {https://doi.org/10.1051/0004-6361/200811170}
  {\bibfield  {journal} {\bibinfo  {journal} {A\&A}\ }\textbf {\bibinfo
  {volume} {508}},\ \bibinfo {pages} {541} (\bibinfo {year}
  {2009})}\BibitemShut {NoStop}%
\bibitem [{\citenamefont {Larsson}\ \emph {et~al.}(2013)\citenamefont
  {Larsson}, \citenamefont {Bermejo-Moreno},\ and\ \citenamefont
  {Lele}}]{larsson2013}%
  \BibitemOpen
  \bibfield  {author} {\bibinfo {author} {\bibfnamefont {J.}~\bibnamefont
  {Larsson}}, \bibinfo {author} {\bibfnamefont {I.}~\bibnamefont
  {Bermejo-Moreno}},\ and\ \bibinfo {author} {\bibfnamefont {S.~K.}\
  \bibnamefont {Lele}},\ }\bibfield  {title} {\bibinfo {title} {Reynolds- and
  mach-number effects in canonical shock–turbulence interaction},\ }\href
  {https://doi.org/10.1017/jfm.2012.573} {\bibfield  {journal} {\bibinfo
  {journal} {Journal of Fluid Mechanics}\ }\textbf {\bibinfo {volume} {717}},\
  \bibinfo {pages} {293–321} (\bibinfo {year} {2013})}\BibitemShut {NoStop}%
\bibitem [{\citenamefont {Federrath}(2013)}]{federrath2013}%
  \BibitemOpen
  \bibfield  {author} {\bibinfo {author} {\bibfnamefont {C.}~\bibnamefont
  {Federrath}},\ }\bibfield  {title} {\bibinfo {title} {On the universality of
  supersonic turbulence},\ }\href {https://doi.org/10.1093/mnras/stt1644}
  {\bibfield  {journal} {\bibinfo  {journal} {Monthly Notices of the Royal
  Astronomical Society}\ }\textbf {\bibinfo {volume} {436}},\ \bibinfo {pages}
  {1245} (\bibinfo {year} {2013})},\ \Eprint
  {https://arxiv.org/abs/http://mnras.oxfordjournals.org/content/436/2/1245.full.pdf+html}
  {http://mnras.oxfordjournals.org/content/436/2/1245.full.pdf+html}
  \BibitemShut {NoStop}%
\bibitem [{\citenamefont {Kim}\ \emph {et~al.}(2021)\citenamefont {Kim},
  \citenamefont {Murphy}, \citenamefont {Kozlowski}, \citenamefont {Green},
  \citenamefont {Haines}, \citenamefont {Day}, \citenamefont {Cardenas},
  \citenamefont {Woods}, \citenamefont {Smidt}, \citenamefont {Douglas},
  \citenamefont {Jones}, \citenamefont {Velechovsky}, \citenamefont {Olson},
  \citenamefont {Gore},\ and\ \citenamefont {Albright}}]{kim2021}%
  \BibitemOpen
  \bibfield  {author} {\bibinfo {author} {\bibfnamefont {Y.}~\bibnamefont
  {Kim}}, \bibinfo {author} {\bibfnamefont {T.~J.}\ \bibnamefont {Murphy}},
  \bibinfo {author} {\bibfnamefont {P.~M.}\ \bibnamefont {Kozlowski}}, \bibinfo
  {author} {\bibfnamefont {L.~M.}\ \bibnamefont {Green}}, \bibinfo {author}
  {\bibfnamefont {B.~M.}\ \bibnamefont {Haines}}, \bibinfo {author}
  {\bibfnamefont {T.~H.}\ \bibnamefont {Day}}, \bibinfo {author} {\bibfnamefont
  {T.}~\bibnamefont {Cardenas}}, \bibinfo {author} {\bibfnamefont {D.~N.}\
  \bibnamefont {Woods}}, \bibinfo {author} {\bibfnamefont {J.~M.}\ \bibnamefont
  {Smidt}}, \bibinfo {author} {\bibfnamefont {M.~R.}\ \bibnamefont {Douglas}},
  \bibinfo {author} {\bibfnamefont {S.}~\bibnamefont {Jones}}, \bibinfo
  {author} {\bibfnamefont {J.}~\bibnamefont {Velechovsky}}, \bibinfo {author}
  {\bibfnamefont {R.~E.}\ \bibnamefont {Olson}}, \bibinfo {author}
  {\bibfnamefont {R.~A.}\ \bibnamefont {Gore}},\ and\ \bibinfo {author}
  {\bibfnamefont {B.~J.}\ \bibnamefont {Albright}},\ }\bibfield  {title}
  {\bibinfo {title} {Experimental validation of shock propagation through a
  foam with engineered macro-pores},\ }\href
  {https://doi.org/10.1063/5.0024697} {\bibfield  {journal} {\bibinfo
  {journal} {Physics of Plasmas}\ }\textbf {\bibinfo {volume} {28}},\ \bibinfo
  {pages} {012702} (\bibinfo {year} {2021})},\ \Eprint
  {https://arxiv.org/abs/https://doi.org/10.1063/5.0024697}
  {https://doi.org/10.1063/5.0024697} \BibitemShut {NoStop}%
\bibitem [{\citenamefont {Landau}\ and\ \citenamefont
  {Lifshitz}(1987)}]{landau1987}%
  \BibitemOpen
  \bibfield  {author} {\bibinfo {author} {\bibfnamefont {L.}~\bibnamefont
  {Landau}}\ and\ \bibinfo {author} {\bibfnamefont {E.}~\bibnamefont
  {Lifshitz}},\ }\bibfield  {title} {\bibinfo {title} {Fluid mechanics},\
  }\href@noop {} {\bibfield  {journal} {\bibinfo  {journal} {Course of
  Theoretical Physics}\ }\textbf {\bibinfo {volume} {6}} (\bibinfo {year}
  {1987})}\BibitemShut {NoStop}%
\bibitem [{\citenamefont {Dhawalikar}\ \emph {et~al.}(2021)\citenamefont
  {Dhawalikar}, \citenamefont {Federrath}, \citenamefont {Davidovits},
  \citenamefont {Teyssier}, \citenamefont {Nagel},\ and\ \citenamefont
  {Remington}}]{dhawalikar2021}%
  \BibitemOpen
  \bibfield  {author} {\bibinfo {author} {\bibfnamefont {S.}~\bibnamefont
  {Dhawalikar}}, \bibinfo {author} {\bibfnamefont {C.}~\bibnamefont
  {Federrath}}, \bibinfo {author} {\bibfnamefont {S.}~\bibnamefont
  {Davidovits}}, \bibinfo {author} {\bibfnamefont {R.}~\bibnamefont
  {Teyssier}}, \bibinfo {author} {\bibfnamefont {S.~R.}\ \bibnamefont
  {Nagel}},\ and\ \bibinfo {author} {\bibfnamefont {B.~A.}\ \bibnamefont
  {Remington}},\ }\bibfield  {title} {\bibinfo {title} {{Driving mode of
  shock-driven turbulence}},\ }\href@noop {} {\bibfield  {journal} {\bibinfo
  {journal} {Monthly Notices of the Royal Astronomical Society}\ } (\bibinfo
  {year} {2021})}\BibitemShut {NoStop}%
\bibitem [{\citenamefont {Barre}\ \emph {et~al.}(1996)\citenamefont {Barre},
  \citenamefont {Alem},\ and\ \citenamefont {Bonnet}}]{barre1996}%
  \BibitemOpen
  \bibfield  {author} {\bibinfo {author} {\bibfnamefont {S.}~\bibnamefont
  {Barre}}, \bibinfo {author} {\bibfnamefont {D.}~\bibnamefont {Alem}},\ and\
  \bibinfo {author} {\bibfnamefont {J.~P.}\ \bibnamefont {Bonnet}},\ }\bibfield
   {title} {\bibinfo {title} {Experimental study of a normal shock/homogeneous
  turbulence interaction},\ }\href {https://doi.org/10.2514/3.13175} {\bibfield
   {journal} {\bibinfo  {journal} {AIAA Journal}\ }\textbf {\bibinfo {volume}
  {34}},\ \bibinfo {pages} {968} (\bibinfo {year} {1996})},\ \Eprint
  {https://arxiv.org/abs/https://doi.org/10.2514/3.13175}
  {https://doi.org/10.2514/3.13175} \BibitemShut {NoStop}%
\bibitem [{\citenamefont {Belan}\ \emph {et~al.}(2010)\citenamefont {Belan},
  \citenamefont {De~Ponte},\ and\ \citenamefont {Tordella}}]{belan2010}%
  \BibitemOpen
  \bibfield  {author} {\bibinfo {author} {\bibfnamefont {M.}~\bibnamefont
  {Belan}}, \bibinfo {author} {\bibfnamefont {S.}~\bibnamefont {De~Ponte}},\
  and\ \bibinfo {author} {\bibfnamefont {D.}~\bibnamefont {Tordella}},\
  }\bibfield  {title} {\bibinfo {title} {Highly underexpanded jets in the
  presence of a density jump between an ambient gas and a jet},\ }\href
  {https://doi.org/10.1103/PhysRevE.82.026303} {\bibfield  {journal} {\bibinfo
  {journal} {Phys. Rev. E}\ }\textbf {\bibinfo {volume} {82}},\ \bibinfo
  {pages} {026303} (\bibinfo {year} {2010})}\BibitemShut {NoStop}%
\bibitem [{\citenamefont {Opher}\ \emph {et~al.}(2021)\citenamefont {Opher},
  \citenamefont {Drake}, \citenamefont {Zank}, \citenamefont {Powell},
  \citenamefont {Shelley}, \citenamefont {Kornbleuth}, \citenamefont
  {Florinski}, \citenamefont {Izmodenov}, \citenamefont {Giacalone},
  \citenamefont {Fuselier}, \citenamefont {Dialynas}, \citenamefont {Loeb},\
  and\ \citenamefont {Richardson}}]{opher2021}%
  \BibitemOpen
  \bibfield  {author} {\bibinfo {author} {\bibfnamefont {M.}~\bibnamefont
  {Opher}}, \bibinfo {author} {\bibfnamefont {J.~F.}\ \bibnamefont {Drake}},
  \bibinfo {author} {\bibfnamefont {G.}~\bibnamefont {Zank}}, \bibinfo {author}
  {\bibfnamefont {E.}~\bibnamefont {Powell}}, \bibinfo {author} {\bibfnamefont
  {W.}~\bibnamefont {Shelley}}, \bibinfo {author} {\bibfnamefont
  {M.}~\bibnamefont {Kornbleuth}}, \bibinfo {author} {\bibfnamefont
  {V.}~\bibnamefont {Florinski}}, \bibinfo {author} {\bibfnamefont
  {V.}~\bibnamefont {Izmodenov}}, \bibinfo {author} {\bibfnamefont
  {J.}~\bibnamefont {Giacalone}}, \bibinfo {author} {\bibfnamefont
  {S.}~\bibnamefont {Fuselier}}, \bibinfo {author} {\bibfnamefont
  {K.}~\bibnamefont {Dialynas}}, \bibinfo {author} {\bibfnamefont
  {A.}~\bibnamefont {Loeb}},\ and\ \bibinfo {author} {\bibfnamefont
  {J.}~\bibnamefont {Richardson}},\ }\bibfield  {title} {\bibinfo {title} {A
  turbulent heliosheath driven by the rayleigh{\textendash}taylor
  instability},\ }\href {https://doi.org/10.3847/1538-4357/ac2d2e} {\bibfield
  {journal} {\bibinfo  {journal} {The Astrophysical Journal}\ }\textbf
  {\bibinfo {volume} {922}},\ \bibinfo {pages} {181} (\bibinfo {year}
  {2021})}\BibitemShut {NoStop}%
\bibitem [{\citenamefont {Inoue}\ \emph {et~al.}(2011)\citenamefont {Inoue},
  \citenamefont {Yamazaki}, \citenamefont {Inutsuka},\ and\ \citenamefont
  {Fukui}}]{inoue2011}%
  \BibitemOpen
  \bibfield  {author} {\bibinfo {author} {\bibfnamefont {T.}~\bibnamefont
  {Inoue}}, \bibinfo {author} {\bibfnamefont {R.}~\bibnamefont {Yamazaki}},
  \bibinfo {author} {\bibfnamefont {S.-i.}\ \bibnamefont {Inutsuka}},\ and\
  \bibinfo {author} {\bibfnamefont {Y.}~\bibnamefont {Fukui}},\ }\bibfield
  {title} {\bibinfo {title} {Toward understanding the cosmic-ray acceleration
  at young supernova remnants interacting with interstellar clouds: Possible
  applications to rx j1713. 7--3946},\ }\href@noop {} {\bibfield  {journal}
  {\bibinfo  {journal} {The Astrophysical Journal}\ }\textbf {\bibinfo {volume}
  {744}},\ \bibinfo {pages} {71} (\bibinfo {year} {2011})}\BibitemShut
  {NoStop}%
\bibitem [{\citenamefont {Barbeau}\ \emph {et~al.}(2021)\citenamefont
  {Barbeau}, \citenamefont {Raman}, \citenamefont {Manuel}, \citenamefont
  {Nagel},\ and\ \citenamefont {Shivamoggi}}]{barbeau2021}%
  \BibitemOpen
  \bibfield  {author} {\bibinfo {author} {\bibfnamefont {Z.}~\bibnamefont
  {Barbeau}}, \bibinfo {author} {\bibfnamefont {K.}~\bibnamefont {Raman}},
  \bibinfo {author} {\bibfnamefont {M.}~\bibnamefont {Manuel}}, \bibinfo
  {author} {\bibfnamefont {S.}~\bibnamefont {Nagel}},\ and\ \bibinfo {author}
  {\bibfnamefont {B.}~\bibnamefont {Shivamoggi}},\ }\bibfield  {title}
  {\bibinfo {title} {{Design of a high energy density experiment to measure the
  suppresion of hydrodynamic instability in an applied magnetic field}},\
  }\href@noop {} {\bibfield  {journal} {\bibinfo  {journal} {Physics of
  Plasmas}\ } (\bibinfo {year} {2021})}\BibitemShut {NoStop}%
\bibitem [{\citenamefont {Brouillette}(2002)}]{brouillette2002}%
  \BibitemOpen
  \bibfield  {author} {\bibinfo {author} {\bibfnamefont {M.}~\bibnamefont
  {Brouillette}},\ }\bibfield  {title} {\bibinfo {title} {The richtmyer-meshkov
  instability},\ }\href
  {https://doi.org/10.1146/annurev.fluid.34.090101.162238} {\bibfield
  {journal} {\bibinfo  {journal} {Annual Review of Fluid Mechanics}\ }\textbf
  {\bibinfo {volume} {34}},\ \bibinfo {pages} {445} (\bibinfo {year} {2002})},\
  \Eprint
  {https://arxiv.org/abs/https://doi.org/10.1146/annurev.fluid.34.090101.162238}
  {https://doi.org/10.1146/annurev.fluid.34.090101.162238} \BibitemShut
  {NoStop}%
\bibitem [{\citenamefont {Lee}\ and\ \citenamefont {More}(1984)}]{lee1984}%
  \BibitemOpen
  \bibfield  {author} {\bibinfo {author} {\bibfnamefont {Y.~T.}\ \bibnamefont
  {Lee}}\ and\ \bibinfo {author} {\bibfnamefont {R.~M.}\ \bibnamefont {More}},\
  }\bibfield  {title} {\bibinfo {title} {An electron conductivity model for
  dense plasmas},\ }\href {https://doi.org/10.1063/1.864744} {\bibfield
  {journal} {\bibinfo  {journal} {The Physics of Fluids}\ }\textbf {\bibinfo
  {volume} {27}},\ \bibinfo {pages} {1273} (\bibinfo {year} {1984})},\ \Eprint
  {https://arxiv.org/abs/https://aip.scitation.org/doi/pdf/10.1063/1.864744}
  {https://aip.scitation.org/doi/pdf/10.1063/1.864744} \BibitemShut {NoStop}%
\bibitem [{\citenamefont {Olson}\ \emph {et~al.}(2006)\citenamefont {Olson},
  \citenamefont {Bradley}, \citenamefont {Rochau}, \citenamefont {Collins},
  \citenamefont {Leeper},\ and\ \citenamefont {Suter}}]{olson2006}%
  \BibitemOpen
  \bibfield  {author} {\bibinfo {author} {\bibfnamefont {R.~E.}\ \bibnamefont
  {Olson}}, \bibinfo {author} {\bibfnamefont {D.~K.}\ \bibnamefont {Bradley}},
  \bibinfo {author} {\bibfnamefont {G.~A.}\ \bibnamefont {Rochau}}, \bibinfo
  {author} {\bibfnamefont {G.~W.}\ \bibnamefont {Collins}}, \bibinfo {author}
  {\bibfnamefont {R.~J.}\ \bibnamefont {Leeper}},\ and\ \bibinfo {author}
  {\bibfnamefont {L.~J.}\ \bibnamefont {Suter}},\ }\bibfield  {title} {\bibinfo
  {title} {Time-resolved characterization of hohlraum radiation temperature via
  interferometer measurement of quartz shock velocity},\ }\href
  {https://doi.org/10.1063/1.2336458} {\bibfield  {journal} {\bibinfo
  {journal} {Review of Scientific Instruments}\ }\textbf {\bibinfo {volume}
  {77}},\ \bibinfo {pages} {10E523} (\bibinfo {year} {2006})},\ \Eprint
  {https://arxiv.org/abs/https://doi.org/10.1063/1.2336458}
  {https://doi.org/10.1063/1.2336458} \BibitemShut {NoStop}%
\bibitem [{\citenamefont {More}\ \emph {et~al.}(1988)\citenamefont {More},
  \citenamefont {Warren}, \citenamefont {Young},\ and\ \citenamefont
  {Zimmerman}}]{more1988}%
  \BibitemOpen
  \bibfield  {author} {\bibinfo {author} {\bibfnamefont {R.~M.}\ \bibnamefont
  {More}}, \bibinfo {author} {\bibfnamefont {K.~H.}\ \bibnamefont {Warren}},
  \bibinfo {author} {\bibfnamefont {D.~A.}\ \bibnamefont {Young}},\ and\
  \bibinfo {author} {\bibfnamefont {G.~B.}\ \bibnamefont {Zimmerman}},\
  }\bibfield  {title} {\bibinfo {title} {A new quotidian equation of state
  (qeos) for hot dense matter},\ }\href {https://doi.org/10.1063/1.866963}
  {\bibfield  {journal} {\bibinfo  {journal} {The Physics of Fluids}\ }\textbf
  {\bibinfo {volume} {31}},\ \bibinfo {pages} {3059} (\bibinfo {year}
  {1988})},\ \Eprint
  {https://arxiv.org/abs/https://aip.scitation.org/doi/pdf/10.1063/1.866963}
  {https://aip.scitation.org/doi/pdf/10.1063/1.866963} \BibitemShut {NoStop}%
\end{thebibliography}
\end{document}